\documentclass[aps,pre,amsmath,amsfonts,floatfix,superscriptaddress,nofootinbib]{revtex4}
    
    \usepackage{amsmath,amssymb,color,bm}
      \usepackage{amsthm}
      \usepackage{graphicx}
     
    \usepackage{ulem}
    \usepackage{overpic}
    \usepackage{mathrsfs} \usepackage{color} \usepackage{multirow}
    \usepackage{bbm}  \usepackage{commath} 
    \usepackage{comment}
    \let\a=\alpha  \let\g=\gamma \let\d=\delta
     \let\z=\zeta  \let\k=\kappa
    \let\l=\lambda \let\m=\mu \let\n=\nu \let\x=\xi \let\p=\pi
     \let\t=\tau \let\f=\varphi \let\c=\chi
       \let\G=\Gamma
    \let\D=\Delta \let\Th=\Theta \let\X=\Xi 
       
    \let\ee=\varepsilon \let\r=\rho \let\th=\theta \let\io=\infty
    
    \def\ie{{\textit{i.e.} }}\def\eg{{\textit{e.g.} }}
    
    \def\PP{{\cal P}}\def\MM{{\cal M}} 
    \def\CC{{\cal C}}\def\FF{{\cal F}} 
     
    \def\cR{{\cal R}}  \def\OO{{\cal O}}
    \def\DD{{\cal D}}\def\GG{{\cal G}} 
    \def\KK{{\cal K}}

    \def\xx{{\bf x}}

      \def\erf{\mathrm{erf}}

    \def\rr{\mathbf{r}}
    
    \def\xx{\mathbf{x}}
    \def\de{\mathrm d}

    \def\erfc{\mathrm{erfc}}

    \newcommand{\bxi}{\boldsymbol{\xi}}

    \newcommand{\blambda}{\boldsymbol{\lambda}}

    \newcommand{\dd}{\text{d}}

    \newcommand{\bx}{\text{\bf x}}
    \newcommand{\br}{\text{\bf r}}

    \newcommand{\bff}{\text{\bf f}}

    \newcommand{\bu}{\text{\bf u}}

    \newcommand{\bnabla}{\boldsymbol{\nabla}}

    \def\to{\rightarrow} \def\la{\left\langle} \def\ra{\right\rangle}

    \newcommand{\beq}{\begin{equation}} \newcommand{\eeq}{\end{equation}}
    \newcommand{\wh}{\widehat}

    \newcommand{\moy}[1]{\left\langle  #1 \right\rangle }

\begin{document}
    
    \title{Active matter in infinite dimensions:  Fokker-Planck equation and dynamical mean-field theory at low density}
    \author{Thibaut Arnoulx de Pirey}
    \affiliation{Universit\'e de Paris, Laboratoire Mati\`ere et Syst\`emes Complexes (MSC), UMR 7057 CNRS, F-75205 Paris, France}
    \author{Alessandro Manacorda}
    \affiliation{Laboratoire de Physique de l'Ecole Normale Sup\'erieure, ENS, Universit\'e PSL, CNRS, Sorbonne Universit\'e, Universit\'e de Paris, F-75005 Paris, France}
    \affiliation{Department of Physics and Materials Science, University of Luxembourg - L-1511 Luxembourg}
    \author{Fr\'ed\'eric van Wijland}
    \affiliation{Universit\'e de Paris, Laboratoire Mati\`ere et Syst\`emes Complexes (MSC), UMR 7057 CNRS, F-75205 Paris, France}
    \author{Francesco Zamponi}
    \affiliation{Laboratoire de Physique de l'Ecole Normale Sup\'erieure, ENS, Universit\'e PSL, CNRS, Sorbonne Universit\'e, Universit\'e de Paris, F-75005 Paris, France}
    
    \begin{abstract}
        We investigate the behavior of self-propelled particles in infinite space 
        dimensions by comparing two powerful approaches in many-body dynamics: 
        the Fokker-Planck equation and dynamical mean-field theory. The dynamics of the particles at low densities and infinite persistence time
        is solved in the steady-state with both methods, thereby proving the consistency 
        of the two approaches in a paradigmatic out-of-equilibrium system. We 
        obtain the analytic expression for the pair distribution function 
        and the effective self-propulsion to first order in the density, confirming
        the results obtained in a previous paper~\cite{ALW19} and 
        extending them to the case of a non-monotonous interaction potential.
        Furthermore, we obtain the transient behavior of active hard spheres 
        when relaxing from equilibrium to the nonequilibrium steady-state. Our 
        results show how collective dynamics is affected by interactions to 
        first order in the density, and point out future directions for further 
        analytical and numerical solutions of this problem.
    \end{abstract}
    
    \maketitle
    
    \section{Introduction}
    
    The dynamics of active systems has become in the last years one of the most 
    fertile research grounds in nonequilibrium statistical physics. This upsurge of interest stems from the emergence of collective behaviors whose phenomenology is deeply rooted in the intrinsic nonequilibrium nature of the dynamics~\cite{VZ12pr,CT15arcmp,BDLLRVV16rmp}. One of the most paradigmatic model of active matter is given by self-propelled particles, \ie 
    particles that are able to move individually without the need of 
    interactions or thermal fluctuations, but rather driven by an internal 
    self-propulsion giving them a characteristic velocity with typical persistence 
    time. Even in absence of attractive or aligning interactions, it has been 
    shown that these models can exhibit spectacular properties such as 
    motility-induced phase separation (MIPS)~\cite{TC08prl} and local polar 
    order~\cite{CMBMP20prl,Henkes20natcomm,Szamel21epl}. Several recent studies also started to investigate 
    the dynamics of dense particle systems where the activity can be continuously 
    increased, in order to assess the effect of activity on the glass 
    transition and jamming~\cite{BK13,FSB16sm,bi2016motility,BFS19jcp}. A theoretical understanding of such 
    systems strongly deals with their many-body nature, especially when 
    considering the dynamics of dense phases. Indeed, in dense active systems, the difficulties and the need of approximations already needed to understand 
    the behavior of equilibrium liquids~\cite{Hansen} combine with the inherently nonequilibrium nature of the dynamics.
  
    The limit of large space dimensionality has gained attraction first in the field of simple liquids 
    for its ability to capture the thermodynamics of dense phases 
    restricting the analytical difficulties to the computation of the 
    second-order virial coefficient~\cite{FRW85}. The large-dimensional limit 
    is a standard tool in statistical mechanics to study phase transitions~\cite{ParisiBook},
    it has been then widely used to study the glass 
    transition~\cite{KW87,CKPUZ17,parisi2020theory}. This framework allowed for 
    the derivation of dynamical mean-field theory (DMFT): the main idea originated in 
    the dynamics of strongly correlated electrons~\cite{GKKR96}, and has been then 
    applied to describe the microscopic, fluctuating 
    dynamics of equilibrium liquids~\cite{MKZ15,Sz17}, and later of general, 
    nonequilibrium dynamics of particle systems~\cite{AMZ19,AMZ19b}. The 
    equilibrium dynamical equations have recently been numerically solved~\cite{MSZ20}, 
    obtaining the force-force correlation kernels~\cite{baity2019mean}
    leading to the identification of a dynamical glass transition for hard and 
    soft spheres interaction potential.
    
    In another recent work,  
    an approximate way to determine the steady-state many-body dynamics of self-propelled
    particles in infinitely many dimensions was presented~\cite{ALW19}. Conversely from the 
    DMFT approach, the dynamics has been described within the framework of kinetic 
    theory (KT), with a proposed closure of the BBGKY hierarchy taking $1/d$ 
    as the small parameter in perturbation theory. Under these assumptions, it 
    has been shown for self-propelled hard spheres how the $n$-particle distribution function is determined by 
    the pair distribution in the infinite-dimensional limit, leading 
    to the derivation of: (i) the effective propulsion of individual 
    particles, \ie the actual speed of a tagged particle subject to 
    active self-propulsion and repulsive interactions; (ii) the equation of state
    for the pressure $P$ of the active system, leading to the observation 
    of MIPS as a spinodal transition when the pressure decreases with the 
    density.
    
    The above-mentioned study represents one of the first investigations 
    on the dynamics of infinite-dimensional particle systems out of 
    equilibrium, in a paradigmatic field such as active matter. In this 
    paper, we compare the two approaches reproducing the results from KT 
    in the DMFT framework, and apply both theories to solve the dynamics 
    of sticky hard spheres at low densities and infinite persistence time 
    - a framework that has been investigated in several recent studies~\cite{stenhammar2014phase,ALW19,Mo20sm,Morse21pnas,Agoritsas21jsm}. We will show how 
    the pair 
    distribution function $g(r)$ and the density-dependent effective
    propulsion $v(\r)$ can be analytically and consistently obtained in both cases. Furthermore, the analytical solution 
    of DMFT equations includes as well the relaxation from a Boltzmann equilibrium state to a nonequilibrium steady state once activity is switched on, 
    giving new insights into the structural properties of these systems.
    
    The paper is organized as follows. Sec.~\ref{sec:def} provides the basic definitions of the model and of its relevant parameters.  In Sec.~\ref{sec:sticky-kinetic}, 
    we derive the expression for the pair distribution function and the effective
    propulsion velocity by means of kinetic theory. In Sec.~\ref{sec:sticky-dmft}, the 
    same results are found by means of the dynamical mean-field theory in the 
    steady state. In Sec.~\ref{sec:transient}, we then reconsider the purely 
    repulsive hard sphere case discussed in~\cite{ALW19}, including the 
    transient effects obtained by DMFT. We then summarize our findings and 
    point out to possible next steps for future investigation.

    \section{Definition of the model}
    \label{sec:def}
    We consider the dynamics of $N$ interacting $d$-dimensional self-propelled particles with equations of motion that read
    \begin{equation} \label{eq:N_body_dyn}
        \zeta \dot{\bx}_i(t) = \bff_i(t) - \sum_{j(\neq i)}\bnabla_{\bx_i}V\left(\bx_i(t) - \bx_j(t)\right) \, .
    \end{equation}
    In Eq.\eqref{eq:N_body_dyn}, motion is induced by (i) a self-propulsion force $\bff_i(t)$ and (ii) pairwise conservative forces deriving from the potential $U\left(\bx_1, \dots \bx_N\right) = \sum_{i < j} V\left(\bx_i - \bx_j\right)$. The pair potential is taken radially symmetric, $V\left(\bx\right) = V(x)$ with $x = \vert \bx \vert$. There exist different descriptions of the driving force, each of them corresponding to a particular model of self-propelled particles. In the following, we choose to work with run-and-tumble particles (RTPs) in which case the active force reads $\bff_i(t) = v_0 \bu_i(t)$ with $\bu_i(t)$ a unit vector randomly and uniformly reshuffled on the $(d-1)$-dimensional unit sphere with rate $\t_p^{-1}$, thus yielding $\moy{u_i^\mu(t)u_j^\nu(s)} = \delta_{ij}\delta^{\mu\nu}\exp{\left(-|t-s|/\t_p\right)}/d$. Note however that, as shown in \cite{ALW19}, the three standard models of self-propelled particles, \textit{i.e.} RTPs, active Brownian particles and active Ornstein-Ulhenbeck particles, are equivalent in the limit where the space dimension $d$ is sent to infinity and the persistence time is large. In the present work, we study the  dynamics Eq.~\eqref{eq:N_body_dyn} at low densities in the limit $d \to \infty$. We follow \cite{AMZ19, AMZ19b} and take the infinite dimensional limit as follows:
    \begin{itemize}
        \item the pair potential is assumed to decay over a short length scale as $V(x) = \wh{V}(h)$ with $h = d(x/\ell - 1)$. The length scale $\ell$ can be viewed as the particle diameter; the rescaled gap $h$ accounts for the interparticle distance in 
        the $d\to\io$ limit. We then have $h<0$ for overlapping particles, and viceversa;
        \item the packing fraction $\varphi$ (or accordingly the number density $\rho$) is such that each particle interacts with $O(d)$ other particles, \textit{i.e.} $\wh{\varphi} = 2^d \varphi /d = \rho \, \Omega_d \ell^d /d^2$ is kept finite with $\Omega_d$ the $d$-dimensional solid angle;
        \item the norm of the active drive is rescaled as $v_0 = (\sqrt{2} d^{3/2}/\ell)\wh{v}_0$. In Eq.~\eqref{eq:N_body_dyn}, this equates the scaling of the norms of the conservative force and of the active force;
        \item the friction coefficient is rescaled as $\zeta = (2 d^2 
        /\ell^2) \wh{\zeta}$. In this scaling, the variations of the rescaled separation $h$ between two particles over finite time scales are $O(1)$;
        \item the times $t$ and $\tau_p$ are left unchanged.
    \end{itemize}
    
    We also remark that in this settings the equilibrium dynamics can 
    be recovered in the limit $\t_p \to 0$ by setting $\wh v_0^2 = \wh\z T / \t_p$. With this prescription, the active force becomes a thermal 
    noise at temperature $T$ in the limit of vanishing persistence.
    
    \section{Sticky spheres}
    
    \subsection{Results from the Fokker-Planck equation}
    \label{sec:sticky-kinetic}

    \subsubsection{The two-body problem in the infinite dimensional limit}
    
    We start by addressing the dynamics of two interacting self-propelled particles. The interaction is carried through a spherically symmetric potential $V$ and both particles are subject to an external active drive. Following Eq.~\eqref{eq:N_body_dyn}, their equations of motion read
    \begin{equation} \begin{split} \label{eq:EOM_RTP}
        \zeta \dot \bx_1(t) & = v_0 \bu_1(t) - \bnabla_{\bx_1}V\left(\bx_1(t) - \bx_2(t)\right) \, , \\
        \zeta \dot \bx_2(t) & = v_0 \bu_2(t) - \bnabla_{\bx_2}V\left(\bx_2(t) - \bx_1(t)\right) \, ,
        \end{split}
    \end{equation}
    with $\bu_1(t)$ and $\bu_2(t)$ two independent run-and-tumble noises. We then introduce $\br = \bx_2 - \bx_1$, the relative separation, and $P(\br, \bu_1, \bu_2)$, the stationary state probability density associated to the process in Eq.~\eqref{eq:EOM_RTP}. The latter obeys the following integro-differential equation
    \begin{equation}
    - v_0 \left(\bu_2 - \bu_1\right) \cdot \bnabla_{\br}P + 2\bnabla_{\br}\cdot\left(P\bnabla_{\br}V(\br)\right) + \frac{\zeta}{\tau_p}\left[\int \frac{\dd \bu'}{\Omega_d}P(\br,\bu',\bu_2) + \int \frac{\dd \bu'}{\Omega_d}P(\br,\bu_1,\bu') - 2P(\br,\bu_1,\bu_2) \right] = 0 \, ,
    \end{equation}
where the terms between brackets account for the tumble dynamics of the active degrees of freedom $\bu_1$ and $\bu_2$. Taking advantage of the rotational symmetry of $P$ we introduce the variables
    \begin{equation} \begin{split} \label{eq:coord}
        r & = \vert \br \vert \, , \\
        w_1 & = \left(\bu_1 \cdot \br \right) / r  \, , \\
        w_2 & = \left(\bu_2 \cdot \br \right) / r \, , \\
        z & = \bu_1 \cdot \bu_2  \, , \\
        \end{split}
    \end{equation}
    the use of which allows us to rewrite to Fokker-Planck equation in terms of four (instead of $3d$) coordinates as
    \begin{equation} \begin{split} \label{eq:FP_coord}
        0 = &- v_0 \left(w_2 - w_1\right)\partial_r P - \frac{v_0}{r} \left[ \left( 1 + w_1 w_2 - z \right)\left(\partial_{w_2} - \partial_{w_1}\right)P - \left( w_2^2 \, \partial_{w_2} - w_1^2 \, \partial_{w_1}\right)P\right] + \frac{2}{r^{d-1}}\partial_r \left( r^{d-1} V'(r) P \right) \\  
        & + \frac{\zeta}{\tau_p}\left[\frac{\Omega_{d-2}}{\Omega_d}\frac{1}{\sqrt{1 - w_2^2}}\int_{-1}^{1} \dd w_1' \int_{w_1'w_2-\sqrt{1-w_2^2}\sqrt{1-w_1'^2}}^{w_1'w_2+\sqrt{1-w_2^2}\sqrt{1-w_1'^2}}\!\!\!\dd z' P(r, w_1', w_2, z') \left(1 - w_1'^2 - \frac{z'^2 + w_1'^2 w_2^2 - 2 z' w_1' w_2}{1 - w_2^2}\right)^{\frac{d-4}{2}} \right. \\ & \left. + \frac{\Omega_{d-2}}{\Omega_d}\frac{1}{\sqrt{1 - w_1^2}}\int_{-1}^{1} \dd w_2' \int_{w_1w_2'-\sqrt{1-w_2'^2}\sqrt{1-w_1^2}}^{w_1w_2'+\sqrt{1-w_2'^2}\sqrt{1-w_1^2}} \!\!\! \dd z' P(r, w_1, w_2', z') \left(1 - w_2'^2 - \frac{z'^2 + w_2'^2 w_1^2 - 2 z' w_2' w_1}{1 - w_1^2}\right)^{\frac{d-4}{2}} \right. \\ & \left. \vphantom{\left(1 - w_2'^2 - \frac{z'^2 + w_2'^2 w_1^2 - 2 z' w_2' w_1}{1 - w_1^2}\right)^{\frac{d-4}{2}}} - 2 P(r, w_1, w_2, z) \right] .
    \end{split}
    \end{equation}
    The limit of infinite dimension $d \to \infty$ is then taken in Eq.~\eqref{eq:FP_coord} with:
    \begin{equation} \begin{split}
        & r = \ell\left( 1 + h/d \right) \, , \\
        & w_1 \rightarrow w_1/\sqrt{d} \, , \\
        & w_2 \rightarrow w_2/\sqrt{d} \, , \\
        & z \rightarrow z/\sqrt{d} \, . \\
    \end{split}
    \end{equation}
    while keeping $h$, and the redefined $w_1$, $w_2$, $z$ fixed.
    The infinite dimensional limit of Eq.~\eqref{eq:FP_coord} is obtained to leading order in $d$ as:
    \begin{equation}\begin{split}\label{eq:FP_dinf}
        & - \frac{\wh{v}_0}{\sqrt{2}}\left(w_2 - w_1\right) \, \partial_h P - \frac{\wh{v}_0}{\sqrt{2}} \left(\partial_{w_2} - \partial_{w_1}\right)P + e^{-h}\partial_h \left( e^{h} \frac{\wh{V}'(h)}{\ell} P \right) \\ &+ \frac{\wh{\zeta}}{\tau_p}\left[\int_{-\infty}^{+\infty} \frac{\dd w_1' \, \dd z'}{2 \pi} \exp{\left(- \frac{w_1'^2}{2}-\frac{z'^2}{2}\right)} P\left(h, w_1', w_2, z'\right) \right. \\ & \left. + \int_{-\infty}^{+\infty} \frac{\dd w_2' \, \dd z'}{2 \pi} \exp{\left(- \frac{w_2'^2}{2}-\frac{z'^2}{2}\right)} P\left(h, w_1, w_2', z'\right) - 2 P\left(h, w_1, w_2, z\right)\right] = 0 \, .
    \end{split}
    \end{equation}

    \subsubsection{Analytical solution with infinite persistence}
\begin{figure}[h!]
\begin{overpic}[width=\textwidth]{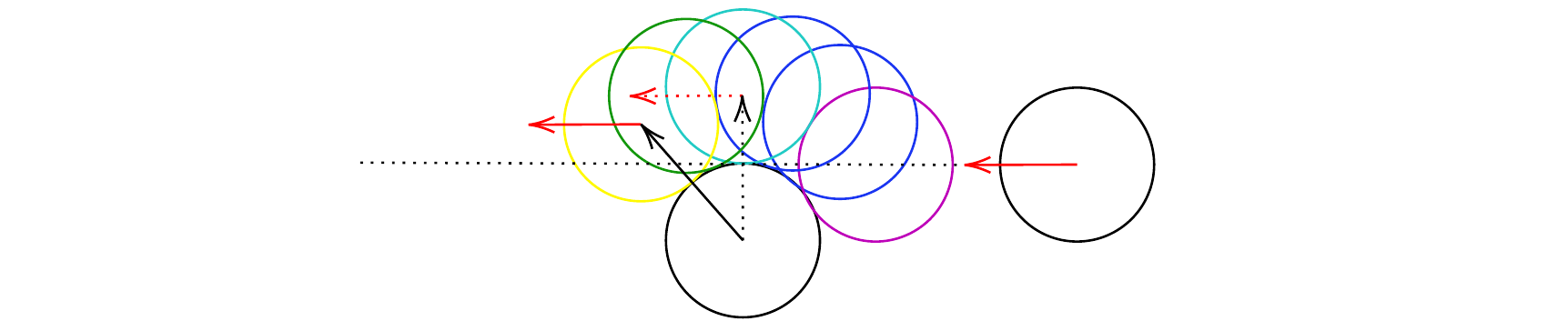}
\put (45,6) {$\hat{\br}$}
\put (48, 5) {1}
\put (69, 9) {2}
\put (65, 11) {\textcolor{red}{$\bu$}}
\put (33, 14) {\textcolor{red}{$\bu$}}
\end{overpic}
\caption{A collision between two active sticky hard spheres labeled 1 and 2 in the reference frame where particle labeled 1 is held fixed. Particle 2 with incoming relative self-propulsion $\bu = \bu_2 - \bu_1$ hits particle 1 (at the magenta position) and then skids around. It eventually takes off at the yellow position where the self-propulsion compensates the attractive interaction between the two spheres, \textit{i.e.} $\hat{\br}\cdot \bu = \sqrt{2}w_0/\sqrt{d}$. The light blue position is where the relative self-propulsion is tangent to the separation between the two spheres and marks the end of the collision in the hard sphere case $w_0 = 0$.}\label{fig:collision}
\end{figure}
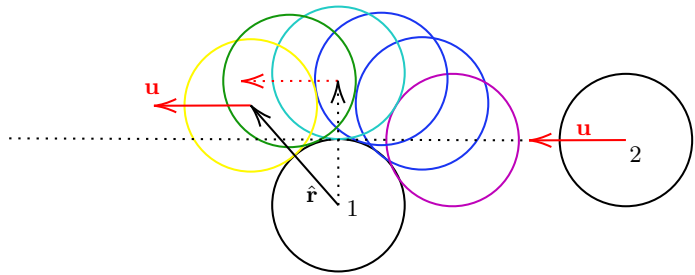

    Equation \eqref{eq:FP_dinf} can be solved analytically for certain classes of potentials in the ballistic limit $\tau_p \to \infty$. Beside providing nice analytical simplifications, this limit is conjectured to be of particular interest regarding the phase behavior of macroscopic systems of interacting active particles, see \cite{stenhammar2014phase} for a discussion in $d = 2$ and $d = 3$ and \cite{ALW19} for a discussion in $d \to \infty$. At $\tau_p \to \infty$, only the relative speed $w = \left(w_2 - w_1\right)/\sqrt{2}$ enters the game and
    \begin{equation} \label{eq:FP_ultraballistic}
        - \wh{v}_0 \left( w \partial_h P + \partial_w P\right) + e^{-h}\partial_h \left( e^{h} \frac{\wh{V}'(h)}{\ell} P \right) = 0
    \end{equation}
    with $P(h \to \infty, w) = 1$ as a boundary condition. Note that similar first order equations also appear in the study of dilute passive colloids at high shear rate \cite{russel1991colloidal}. The class of potentials we work with in the following is that of sticky-sphere potentials. These potentials have hard-sphere repulsion at $h < 0$ while displaying an infinitely short ranged attractive well at $h = 0^+$ and are vanishing at $h > 0$. Such potentials are similar in spirit to the Baxter potential sometimes used as a model for passive colloids with short ranged attraction \cite{baxter1968percus}. However, we make use of a slightly different mathematical construction of these sticky-sphere potentials. Indeed, the pairwise force, when attractive, must always be finite for the stationary state to be well-defined. Were this not to be true, then the two particles whose dynamics is given in Eq.~\eqref{eq:EOM_RTP} would never separate after a collision, the driving forces being unable to counterbalance the attractive force created by the potential. The sticky sphere potential is constructed as follows:
    \beq\label{eq:sticky_pot}
    \wh{V} (h) = \begin{cases}
    	\wh{v}_0 \, w_0 \left( \frac{\l}2 h^2 + h - \frac1{2\l} \right) \ , &\quad h<0 \ , \\
    		\wh{v}_0 \, w_0 \left( - \frac{\l}2 h^2 + h - \frac1{2\l} \right) \ , &\quad 0<h<1/\l \ ,\\
    	0 \ , &\quad h> 1/\l \ ,
    \end{cases}
    \eeq
    in the limit $\lambda \to \infty$, where $w_0$ is a real positive parameter and $2 \wh{v}_0 w_0$ is the maximal attractive force between two particles. The results shown in the following are however independent of the precise procedure used to construct the sticky-sphere potential as the limit of a regular one. Concretely, when colliding, the two spheres skid one onto each other until they are free to go. In the hard-sphere case, this occurs whenever the relative driving force is orthogonal to the relative separation, \textit{i.e.} at $w = 0$. In the sticky-sphere case, they keep skidding and only detach at $w = w_0$, when the projection of the relative driving on the separation direction $2 \wh{v}_0 \, w$ compensates the maximal attractive force, as depicted in Fig.~\ref{fig:collision}
As shown in Appendix~\ref{app:FP}, in the limit $\lambda \to \infty$, the stationary probability distribution splits into a bulk part at $h > 0$ and a delta peak accumulation at $h = 0$,
    \begin{align} \label{eq:stat_prob}
        P(h, w) = P_b(h, w)\Theta(h) + \Gamma(w) \delta(h) \, ,
    \end{align}
   where $\Theta(h)$ is the Heaviside step function and $\delta(h)$ is a Dirac delta, with 
    \begin{align} \label{eq:bulkFP}
        w \partial_h P_b + \partial_w P_b = 0 \, ,
    \end{align}
    and 
    \begin{align} \label{eq:boundaryFP}
       \Gamma'(w) - w \Gamma(w) = - w P_b(0, w) \, \, \text{with} \, \, \Gamma(w > w_0) = 0 \, .
    \end{align}
    The stationary distribution function is shown to be given by (see Appendix~\ref{app:FP} for details of the derivation)
    \begin{align} \label{eq:bulk}
        P_b(h,w) = \Theta(h)\left[1 - \Theta(w)\Theta\left(\frac{w^2}{2}-h\right) + \Theta(w)e^{\frac{w_0^2}{2}}\delta\left(h - \frac{w^2}{2} + \frac{w_0^2}{2}\right)\right]
    \end{align}
    and 
    \begin{align}\label{eq:surface}
        \Gamma(w) = \Theta(-w) + \Theta(w)\Theta(w_0-w)e^{\frac{w^2}{2}} \ .
    \end{align}
    As discussed in the Appendix~\ref{app:FP}, the $2h - w^2 = \text{cst}$ parabolas correspond to the deterministic trajectories (excluding collision events) in the $h,w$ plane. In this plane, the  $\{h > 0, \, w > 0 \, , w^2 - 2h >0 \}$ domain is made of trajectories emanating from a collision event. Equation \eqref{eq:bulk} thus states that the probability to find the system in this region is concentrated on the $w^2 - 2h = w_0^2$ branch: all trajectories with a collision collapse on this line when the two particles detach. As $w_0 \to 0$ (this limit being taken after the $\lambda \to \infty$ one), one obtains the ballistic limit of the stationary probability distribution of two active hard spheres. The marginal in space probability distribution can then be obtained from equations \eqref{eq:bulk}-\eqref{eq:surface} as,
    \begin{equation}
        \begin{split} \label{eq:space}
            P(h) & = \int_{-\infty}^{+\infty}\frac{\dd w_1 \dd w_2}{2\pi} \exp{\left(-\frac{w_1^2}{2}-\frac{w_2^2}{2}\right)}P\left(h, w = \frac{w_2 - w_1}{\sqrt{2}}\right) =  \int_{-\infty}^{+\infty}\frac{\dd w}{\sqrt{2\pi}} e^{-\frac{w^2}{2}} P(h,w)  \\
            & = \Theta(h)\left[\frac{1}{2}\left(1 + \erf\left(\sqrt{h}\right)\right) + \frac{e^{-h}}{\sqrt{2\pi\left(2h + w_0^2\right)}}\right] + \left(\frac{1}{2}+\frac{w_0}{\sqrt{2\pi}}\right)\delta(h) \, .
        \end{split}
    \end{equation}
    The distribution in Eq.~\eqref{eq:space} clearly shows an activity induced attraction between the two particles. The $w_0$ parameter  of the sticky sphere potential controls the amplitude of the attractive delta peak at contact.
    \subsubsection{Thermodynamic properties in the dilute limit}
    We return to the above mentioned general $N$-body dynamics Eq.~\eqref{eq:N_body_dyn}. Deriving the macroscopic properties of the system, such as its two-point function, directly from the set of equations in Eq.~\eqref{eq:N_body_dyn} is in general a formidable task. Here we use the results obtained above to describe the thermodynamic properties of the stationary state of the process in Eq.~\eqref{eq:N_body_dyn} in the dilute limit. In the limits $\wh{\varphi} \to 0$ and $\tau_p \to \infty$, the two point function of the system is given by that of the two-particle one,
    \begin{align} \label{eq:two_point}
        g\left(\br, \bu_1 ; \br', \bu_2\right) = P(h, w) \, ,
    \end{align}
    where the distribution $P$ was previously derived in Eq.~\eqref{eq:stat_prob} with $h = d(|\br - \br'|/\ell - 1)$ and $w = \sqrt{d} \, (\bu_2 - \bu_1)\cdot (\br' - \br)/(\sqrt{2}|\br - \br'|)$ . From Eq.~\eqref{eq:two_point}, we compute two important quantities, a dynamical and a thermodynamical one: the effective self-propulsion $v(\wh{\varphi})$ and the mechanical pressure $p(\wh{\varphi})$. The former gives the average value of the velocity of a single tagged particle conditioned on its self propulsion and is obtained from
    \begin{align} \label{eq:def_veff}
        \zeta \left\langle \dot{\bx}_i(t) \right\rangle_{\bu_i} = v(\varphi)\bu_i \ ,
    \end{align}
    whereas the latter gives information about the spinodal instability of homogeneous phases and phase separation in active systems~\cite{solon2018generalized}. The spinodal line, that signals the onset of linear instability of homogeneous phases, is indeed found through the condition $p'(\wh{\varphi})=0$. From Eq.~\eqref{eq:def_veff}, the effective self-propulsion writes at first order in $\wh{\varphi}$
    \begin{equation}
    \begin{split} \label{eq:veff}
    v(\wh{\varphi}) & = v_0 + \rho \int \dd \br \frac{\dd \bu'}{\Omega_d}g(\mathbf{0},\bu;\br,\bu') V'(r) \hat{\br}\cdot \bu  \\ & = v_0 + d^{3/2}\wh{\varphi} \int \dd h \, e^h \, \frac{\dd w_1 \dd w_2}{2\pi} \exp{\left(-\frac{w_1^2}{2}-\frac{w_2^2}{2}\right)}P\left(h , w = \frac{w_2 - w_1}{\sqrt{2}}\right) \frac{\bar{V}'(h)}{\ell}w_1  \\ & = v_0 \left( 1 + \frac{\wh{\varphi}}{\sqrt{2}} \int \, \frac{\dd w_1 \dd w_2}{2\pi} \exp{\left(-\frac{w_1^2}{2}-\frac{w_2^2}{2}\right)}\Gamma\left(w = \frac{w_2 - w_1}{\sqrt{2}}\right)\frac{w_2 - w_1}{\sqrt{2}} \, w_1 \right)  \\ & = v_0 \left( 1 -  \frac{\wh{\varphi}}{2} \int \, \frac{\dd w}{\sqrt{2\pi}} \exp{\left(-\frac{w^2}{2}\right)}\Gamma\left(w \right)w^2 \right)  \\
    & = v_0\left(1 - \frac{\wh{\varphi}}{4}\left(1 + \frac{\sqrt2 \, w_0^3}{3\sqrt{\pi}}\right)\right) \, ,
    \end{split}
    \end{equation}
    from which it appears clearly that at small density the slow-down of the effective self-propulsion induced by collisions increases with the stickiness of the potential. In order to go from the second to the third line of Eq.~\eqref{eq:veff}, we have used the regularization of the product $P \wh{V}'(h)$ in the hard $\lambda \to \infty$ limit:
    \begin{equation} \label{eq:regular}
       \lim_{\lambda \to \infty} P(h,w) \frac{\wh{V}'(h)}{\ell} = \wh{v}_0 w \, \Gamma(w) \, \delta(h) \ .
    \end{equation}
    A proof of Eq.~\eqref{eq:regular} is given in Appendix~\ref{app:FP}. Next we compute the equation of state for the mechanical pressure associated to Eq.~\eqref{eq:N_body_dyn}. The general expression reads
    \begin{equation}
        p(\wh{\varphi}) = \rho \frac{v_0^2 \tau_p}{d \zeta}\frac{v(\wh{\varphi})}{v_0} - \frac{\rho^2}{2d}\int \dd \br \frac{\dd \bu_1}{\Omega_d}\frac{\dd \bu_2}{\Omega_d}g\left(0,\bu ; \br, \bu'\right) V'(r) r \,.
    \end{equation}
    Furthermore, within the considered scalings,
    \begin{equation} \begin{split}
        \frac{\rho^2}{2d}\int \dd \br \frac{\dd \bu_1}{\Omega_d}\frac{\dd \bu_2}{\Omega_d}g\left(0,\bu ; \br, \bu'\right) V'(r) r & = d \rho \frac{\wh{\varphi}}{2} \ell \int \dd h \, e^h \, \frac{\dd w_1}{\sqrt{2\pi}}\frac{\dd w_2}{\sqrt{2\pi}} \exp{\left(-\frac{w_1^2}{2}-\frac{w_2^2}{2}\right)}g\left(h, w = \frac{w_2 - w_1}{\sqrt{2}}\right) \frac{\bar{V}'(h)}{l}  \\ & = d \rho \frac{\wh{\varphi}}{2} \ell \wh{v}_0 \int \frac{\dd w_1}{\sqrt{2\pi}}\frac{\dd w_2}{\sqrt{2\pi}} \exp{\left(-\frac{w_1^2}{2}-\frac{w_2^2}{2}\right)}\Gamma\left(w = \frac{w_2 - w_1}{\sqrt{2}}\right)w \, \\ & = d \rho \frac{\wh{\varphi}}{2} \ell \wh{v}_0\int \frac{\dd w}{\sqrt{2\pi}} \exp{\left(-\frac{w^2}{2}\right)}\Gamma\left(w\right)w \, \\ & = - d \rho \frac{\wh{\varphi}}{4} \frac{\sqrt{2}\ell \wh{v}_0}{\sqrt{\pi}}\left(1 - \frac{w_0^2}{2}\right) \ .
    \end{split} \end{equation}
    Thus, up to second order in $\varphi$, we obtain the equation of state for the mechanical pressure as
    \begin{equation}\label{eq:p}
    \begin{split}
        \left(\frac{\Omega_d \ell^d}{d^2} \right) \frac{p(\wh{\varphi})}{d}= \wh{\varphi} \frac{\wh{v}_0^2 \wh{\tau}_p}{\wh{\zeta}}\left[1 - \frac{\wh{\varphi}}{4}\left(1 + \frac{\sqrt2 \, w_0^3}{3\sqrt{\pi}}\right)\right] + \frac{\wh{\varphi}^2}{4}\frac{\sqrt{2}\ell\wh{v}_0}{\sqrt{\pi}}\left(1 - \frac{w_0^2}{2}\right) \ .
    \end{split}
    \end{equation}
    Note that in order for the two terms in the above expression to have the same scaling in $d$, we had to rescale the persistence time consistently with the ballistic limit as $\tau_p = d \wh{\tau}_p$. For $\tau_p$ = O(1), the equation of state is dominated by the second, equilibrium-like, term. Note also the manifestly destabilizing role of the sticky-sphere parameter $w_0$ on the homogeneous state.
    \begin{figure}[!ht]
        \centering
        \includegraphics[width=0.47\textwidth]{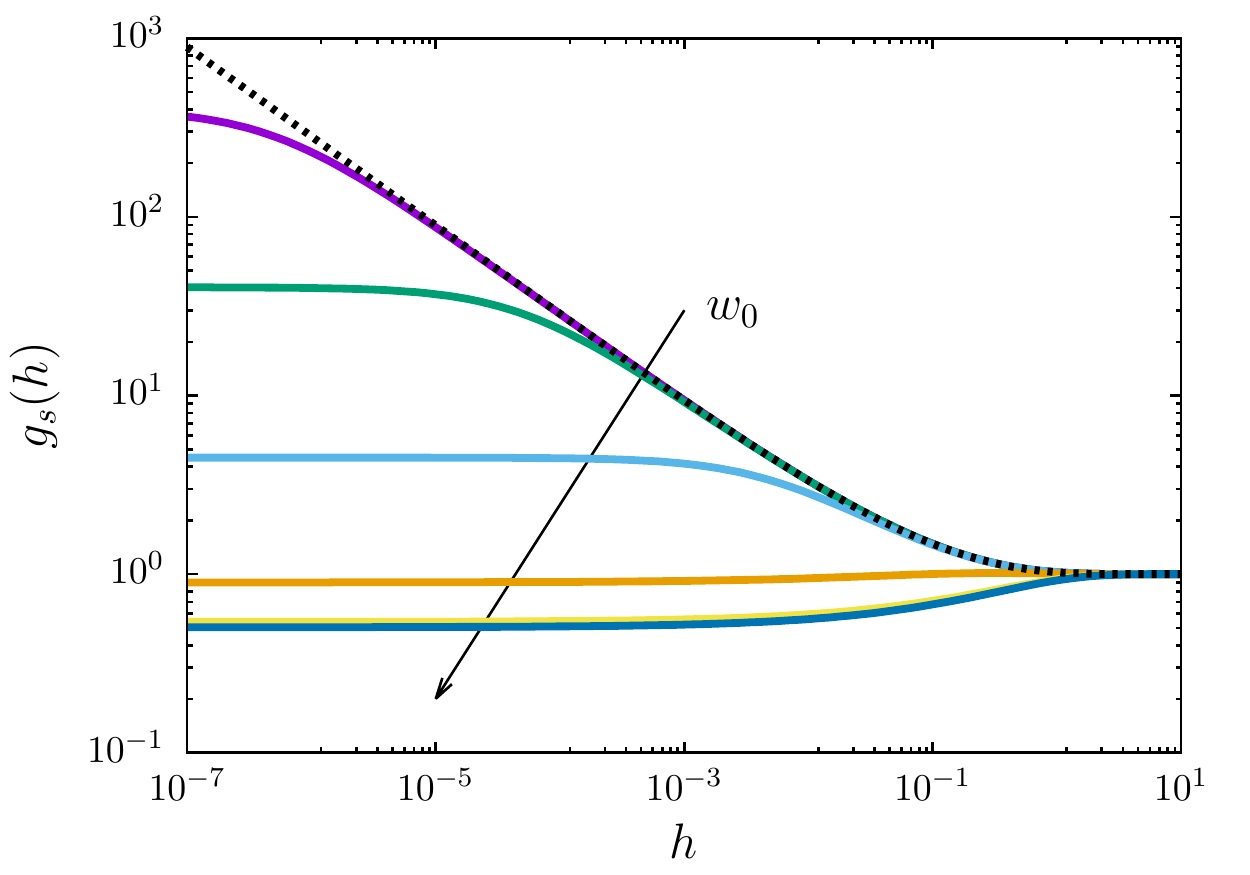}
        ~
        \includegraphics[width=0.47\textwidth]{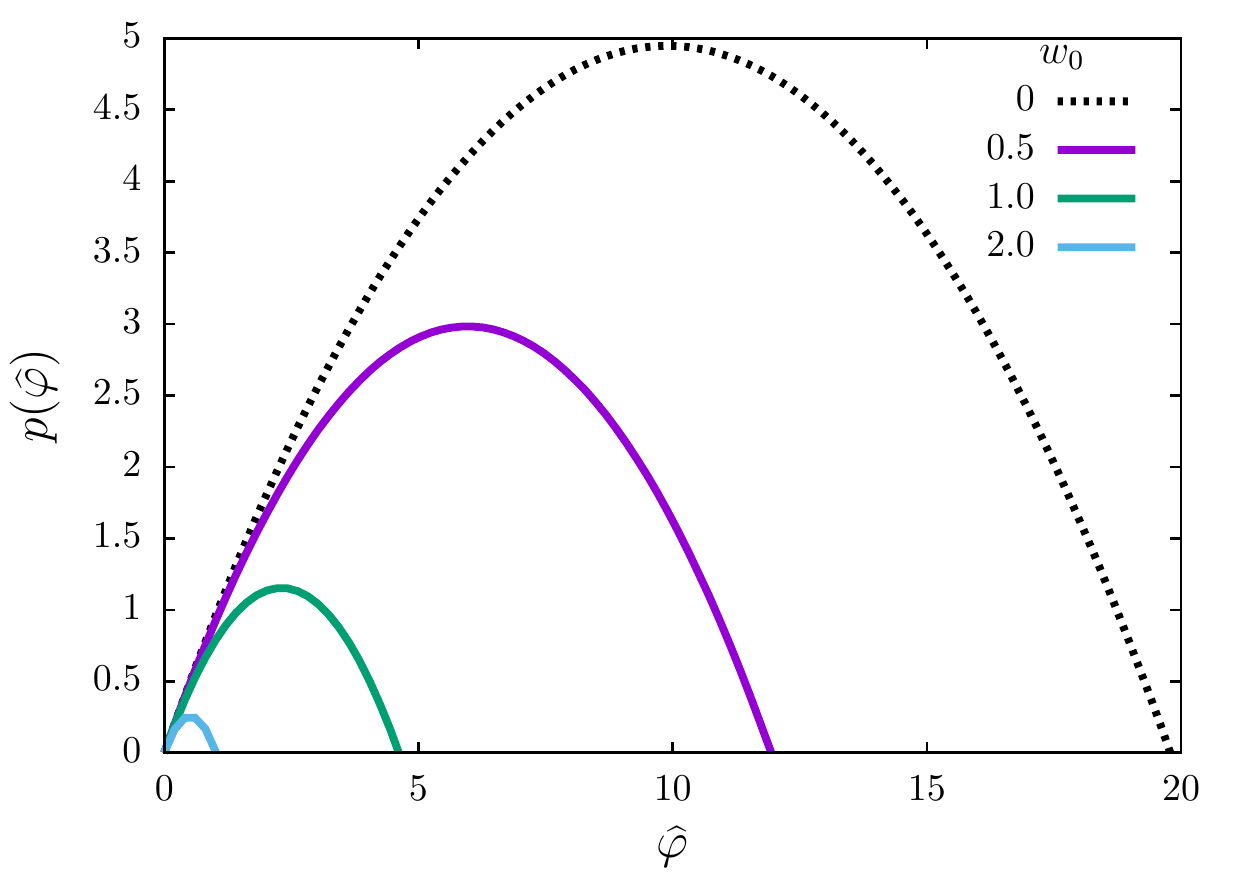}
        \caption{Left: pair distribution function $g_s(h)=P(h)$ vs $h$ in 
        the steady state for $w_0=0$ (purely repulsive case, black dashed line) and $w_0=10^{-3},10^{-2},\ldots,10^2$ (colored lines), 
        Eq.~\eqref{eq:space}.
        The repulsive case displays the $h^{-1/2}$ divergence, while 
        the attractive $w_0>0$ curves have a finite limit at $h=0$; the 
        attractive force monotonically depletes the small $h$ region 
        favoring adhesion at $h=0$, as shown by the delta peak amplitude
        increasing with $w_0$.
        Right: Pressure vs rescaled density as from Eq.~\eqref{eq:p}, with
        $\wh\z=\wh\t_p=\wh v_0=\ell=1$.
        Its behavior is non-monotonic and the decreasing region 
        $\de p/\de \r <0$ is a possible signal of motility-induced phase separation. The pressure becomes negative 
        after a threshold value of $\wh\f$, signaling the unphysical
        behavior of the computed result.}
        \label{fig:p}
    \end{figure}

    \subsection{Results from dynamical mean-field theory}
    \label{sec:sticky-dmft}
    
    \subsubsection{Microscopic dynamics and infinite-dimensional limit}
    
The general DMFT of infinite-dimensional particle systems 
    interacting through pair potentials and subject to external drivings 
    has been derived in~\cite{AMZ19}. Here we  address the dynamics of active 
    particles introduced in Eq.~\eqref{eq:N_body_dyn}, considering the 
    case of an active Ornstein-Uhlenbeck self-propulsion detailed therein.
    The latter is microscopically different from run-and-tumble 
    self-propulsion; nevertheless, we recall that the two active forces 
    are equivalent in the limit of infinite space dimension and 
    persistence time. We choose therefore the Ornstein-Uhlenbeck 
    self-propulsion because of its Gaussianity, consistent with the DMFT 
    derivation in~\cite{AMZ19}.
    The dimensional scaling of self-propulsion, friction coefficient and density 
    follows the prescriptions introduced in Sec.~\ref{sec:def}.
    
    The DMFT framework allows one to describe the $N$-body, $d$-dimensional process in Eq.~\eqref{eq:N_body_dyn} by means of a two-body scalar process; indeed, it is known that when $d \to \io$ the two-particle process can be determined self-consistently by analyzing the behavior of the rescaled inter-particle gap, \ie
    \beq
    h(t) = h_0 + y(t) + \D_r(t) \approx d \left( \frac{r(t)}{\ell} - 1 \right) ,
    \eeq
    where $r(t)$ is the relative distance between two reference particles, 
    $y(t) = (d/\ell) \, \hat \rr_0 \cdot \left( \rr(t) - \rr_0 \right)$ is 
    the rescaled projection of the relative displacement along the initial 
    relative direction, and $\D_r(t) = (d/\ell^2) \left\langle \vert 
    \xx(t) - \xx_0 \vert^2 \right\rangle $ is the mean-square displacement (MSD) contribution given 
    by the $d-1$ transverse components, which is equivalent to the single-particle MSD in 
    the $d\to\io$ limit~\cite{AMZ19}. The equation of motion for $y(t)$ can be shown to take the following form:
    \beq\label{eq:y}
    \begin{split}
     &  \wh\z \dot y(t)
     	=	- \k(t) y(t)
     		+ \int_0^t \!\! \de s \, \MM_R(t,s) \, y(s)
     		- \wh V '(h_0+y(t)+\D_r(t)) +  \X(t)
     \ , \qquad y(0)=0 \ ,\\
     & \moy{ \X(t)}=0
      	\ , \quad
     	\moy{ \X(t) \X(s)}
     		= 2 \wh\z T \, \d(t-s) +  \GG_C(t-s)+ \MM_C(t,s) \ , \qquad
    \GG_C(t)=\wh v_0^2 e^{-|t|/\t_p} 
     \ . \\
    \end{split}
    \eeq
    The colored noise $\X(t)$ has three contributions: (i) the equilibrium 
    thermal bath at temperature $T$. We will drop this 
    term since we will consider the athermal case $T=0$ in the following, 
    but we include it now for the sake of generality; (ii) the active 
    self-propulsion with stationary time correlations $\GG_C(t-s)$, 
    corresponding to active Ornstein-Uhlenbeck particles; (iii) 
    the kernel $\MM_C(t,s)$, accounting for the 
    force-force correlation given by pairwise interactions. The term
    $-\wh V'(h(t))$ is the rescaled two-particle interaction force.
    Finally, DMFT also introduces the instantaneous and retarded response 
    kernels, respectively $\k(t)$ and $\MM_R(t,s)$, to describe the 
    reaction of the $N$-body system on the two-particle process.\\
    The response and correlation kernels $\k(t)$, $\MM_R(t,s)$ and $\MM_C(t,s)$ need to be determined self-consistently with the definitions
    \beq\label{eq:ker}
    \begin{split}
     \k(t)
     	&= \frac{\wh \f}2 \int^{\infty}_{-\infty} \, \de h_0 \, e^{h_0} g_0(h_0)   \la \wh V''(h(t)) + \wh V'(h(t)) \ra_{h_0}
     \ , \\
     \MM_C(t,t')
     	&=  \frac{\wh\f}2 \int^{\infty}_{-\infty}  \de h_0 \, e^{h_0} g_0(h_0)  
     		\la \wh V'(h(t)) \wh V'(h(t')) \ra_{h_0} 
     \ , \\
     \MM_R(t,t')
     	&=  \frac{\wh\f}2 \int^{\infty}_{-\infty}  \de h_0 \, e^{h_0} g_0(h_0)  
    		\left. \frac{\d \la \wh V'(h(t))  \ra_{h_0,\PP}}{\d \PP(t')}\right\vert_{\PP=0} \\
    		&= \frac{\wh\f}2 \int^{\infty}_{-\infty}  \de h_0 \, e^{h_0} g_0(h_0)  
    		\la \wh V''(h(t)) H(t,s) \ra_{h_0}
     \ ,
    \end{split}
    \eeq
    where $g_0(h_0)$ is the initial gap distribution function, $\moy{\ldots}_{h_0}$ refers to 
    an average over the trajectory realizations conditioned to the initial condition 
    $h(0)=h_0$, the 
    perturbation ${\PP(t)}$ acts in the pairwise interaction as
    ${\wh V'(h_0+y(t)+\D_r(t)) \to \wh V'(h_0+y(t)+\D_r(t) - \PP(t))}$, 
    and the fluctuating response is defined as
    $H(t,s) = \d h(t) /\d \PP(s) \vert_{\PP=0}$; its evolution is given by
     \beq\label{eq:H}
     \wh\z \frac{\partial}{\partial t} H(t,t') = - \k(t) H(t,t') - 
     \wh V''(h(t)) \left[ H(t,t') - \d(t-t') \right] + \int^t_{t'} \de s \,
     \MM_R(t,s) H(s,t') \ .
     \eeq 
     The system is not yet closed, because of the MSD contribution given 
     by $\D_r(t)$ in Eq.~\eqref{eq:y}; the latter can be determined 
     through the one-particle dynamical correlation and response defined 
     in~\cite{AMZ19} as
     \beq \label{eq:response_correl}
     \CC(t,t') = \frac{d}{N \ell^2} \sum_{i=1}^N \moy{ \d \xx_i (t) \cdot \d \xx_i(t') } \ , \quad 
     \cR(t,t') = \frac{d}{N \ell^2} \sum_{i,\m} \left. \frac{\d \moy{\d x_{i\m}(t)}}{\d \l_{i\m}(t') } \right\vert_{\blambda=0} \ ,
     \eeq
     where $\d \xx_i (t) = \xx_i(t) - \xx_i(0)$ is the relative displacement of particle $i$ with respect to its initial position and where the perturbation $\blambda_i$ appearing in the definition of the response function $\cR(t,t')$ acts at the one-particle level as
     \begin{equation}
     \zeta \dot{\bx}_i(t) = \bff_i(t) - \sum_{j (\neq i)} \bnabla_{\bx_i}V(\bx_i - \bx_j) + \blambda_i(t) \, .
     \end{equation}
     In the limit of infinite dimension, the correlation and response evolve according to the following dynamics
    \begin{equation}\label{eq:corr}
    \begin{split}
    \wh \z \frac{\partial}{\partial t} \CC(t,t')
      	=& 2 \wh\z T \cR(t',t) - \k(t)\CC(t,t')+\int_0^{t}\de s\,\MM_R(t,s)\CC(s,t')
     		 +\int_0^{t'}\de s\,\left[\GG_C(t-s)+\MM_C(t,s)\right]\cR(t',s)
     \ ,\\
    \wh \z \frac{\partial}{\partial t}
     \cR(t,t')
     	=& \frac{\d(t-t')}{2}-\k(t)\cR(t,t')+\int_{t'}^{t}\de s\,\MM_R(t,s)\cR(s,t')
     \ .
     \end{split}
    \end{equation}
    By definition, one has $\D_r(t) = \CC(t,t)$, and the dynamical 
    equations are at this stage closed. The evolution equation for 
    the MSD $\D(t,t') = \frac{d}{\ell^2} \moy{\vert \xx(t)-\xx(t') 
    \vert^2}$ and $\D_r(t) \equiv \D(t,0)$ therefore read
    \beq\label{eq:MSD}
    \begin{split}
        \wh \z \frac{\partial}{\partial t} \D(t,t')
      	=&-\k(t) \left[\D(t,t')+\D_r(t)-\D_r(t')\right] + 
        \int_0^{t} \de s \, \MM_R(t,s) 
      	\left[\D_r(t)-\D_r(t')+\D(s,t')-\D(s,t)\right] \\
      	& - 4 \wh\z T \, \cR(t',t) + 2\int_0^{\max(t,t')}\de s \, \left[\GG_C(t-s)+\MM_C(t,s)\right] 
      	\left[\cR(t,s)-\cR(t',s)\right] \ , \\
      	\wh\z \dot \D_r(t) =& - 2 \k(t)\D_r(t) + \int_0^{t} \de s \, 
      	\MM_R(t,s) \left[\D_r(t)+\D_r(s)-\D(s,t)\right] + 2 \int_0^t \de s
      	\, \left[\GG_C(t-s)+\MM_C(t,s)\right] \cR(t,s) \ .
    \end{split}
    \eeq
    We stress that, in this framework, the solution of the dynamics is not stationary nor
    time-translationally invariant and depend on the initial condition, \ie the 
    choice of the initial distribution $g_0(h_0)$. The solution of the DMFT equations 
    therefore yields the transient dynamics at short times and the eventual steady-state 
    dynamics at long times.

    \subsubsection{Dilute solution with infinite persistence}
    
    The analytical solution of the problem determined by 
    Eqs.~(\ref{eq:y}-\ref{eq:corr}) is currently out of reach. 
    In the equilibrium case, these equations simplify thanks 
    to fluctuation-dissipation relations and 
    a numerical solution has been found~\cite{MSZ20}. 
    In the present case, a numerical solution must deal with strong technical 
    difficulties, the main one being the sampling efficiency at long times:
    indeed, particles with infinite persistence time eventually collide 
    with a rate that is exponentially decaying in time. Therefore, the 
    amount of trajectories needed to compute the dynamical kernels at long
    time is exponentially high. A possible solution may involve the 
    generation of biased trajectories to increase efficiency, but its 
    design goes beyond the scope of this article.
    
    It is however possible to derive an analytical solution in the 
    dilute limit: indeed, the implicit equations~\eqref{eq:ker} for the 
    kernels depend on the density only through a global multiplicative coefficient.
    Therefore, a solution for \eg the instantaneous response $\k(t)$ reads
    \beq\label{eq:k-iterative}
    \k(t) = \wh\f \, \FF \left[ \k, \MM_R, \MM_C \right] (t) \ .
    \eeq
    An iterative solution can be found assuming that the low-density limit 
    is continuous and that the series
    \beq\label{eq:k-exp}
    \k(t) = \wh\f \, \k^{(1)} (t) + \wh\f^2 \k^{(2)}(t) + \ldots \ .
    \eeq
    converges.  An iterative solution therefore starts with a guess on the initial 
    kernels $\k(t), \: \MM_R(t,t')$ and $\MM_C(t,t')$;  after
    solving the stochastic dynamics in Eqs.~(\ref{eq:y},\ref{eq:H}) with 
    fixed kernels, the latter are updated through Eq.~\eqref{eq:ker}. The 
    self-consistent kernels are given by the fixed points of 
    Eq.~\eqref{eq:k-iterative}.
    
    When $\wh\f=0$, the kernels are trivially vanishing because no 
    interaction occurs. In the dilute limit $\wh\f \ll 1$, the solution 
    can be approximated by the first-order expansion in 
    Eq.~\eqref{eq:k-exp}. The latter can be analytically computed in the 
    infinite persistence time limit $\t_p \to \io$: indeed, in that case 
    the active force reduces to a constant driving and, in absence of 
    dynamical kernels, the trajectories in Eq.~\eqref{eq:y} are 
    fully determined by the self-propulsion $\X(t) \equiv \X_0$ drawn 
    at $t=0$.
    
    The solution of the fluctuating equations~\eqref{eq:y} 
    and~\eqref{eq:H} can be then computed imposing 
    $\k(t)=\MM_R(t,t')=\MM_C(t,t')=0$ and plugging the trajectories 
    $h(t)$ into Eqs.~\eqref{eq:ker} to compute the first-order kernels.
    
    \subsubsection{Analytical solution in the dilute limit: trajectories 
    and pair distribution function}
    \label{sec:DMFT-g}
    
    In the case of vanishing kernels and at $T=0$, the response and 
    correlation read
    \beq\label{eq:CR-1}
    \begin{split}
    \cR(t,t') &= \frac1{2\wh\z} \, \th(t-t') \ , \\
    \CC(t,t') &= \frac{\wh v_0^2}{2\wh\z^2} \, t \,t' \quad \Rightarrow 
    \quad \D(t,t') = \frac{\wh v_0^2}{2\wh\z^2} (t-t')^2 \ .
    \end{split}
    \eeq
    This solution is nothing but the dynamics of a single free active particle moving 
    across a medium with rescaled friction coefficient $\wh\z$, rescaled self-propulsion
    $\wh v_0$ and infinite persistence time $\t_p$. Indeed the response is a step 
    function, \ie a perturbation of the position at time $t'$ remains unchanged at any $t>t'$, 
    and the MSD makes clear that the particle moves ballistically with effective speed 
    $\wh v_0 / \wh\z$. This result represents the first step towards a two-particle solution, 
    and depends on a natural time scale $\t_0 = \wh\z/\wh v_0$, which  
    represents the typical duration of a collision, as it is the time needed 
    to traverse a distance $\ell/d$ at speed $v_0$. This time scale
    must not be confused with $\t_p$, which we recall to be the persistence time 
    of the active self-propulsion. In the 
    following, we will set $\wh\z = \wh v_0 = 1$, setting $\t_0$ as unit 
    of time and $\wh v_0$ as unit of energy; the dimensional coefficients 
    will be reinstated in the final results.
    The solution in Eq.~\eqref{eq:CR-1} 
    leads to the dynamical equation for $h(t) = h_0 + y(t) + \D_r(t)$
    \beq\label{eq:h-1}
    \begin{gathered}
    \dot h(t) = - \wh V'(h(t)) + \x_0 + t \ , \quad h(0) = h_0 \\
    \moy{\x_0} = 0 \ , \quad \langle \x_0^2 \rangle = 1 \ ,
    \end{gathered} 
    \eeq
    having now called $\X(t) = \X_0 = \wh v_0 \, \x_0$. The equation for 
    the fluctuating response $H(t,t')$ now reads
    \beq\label{eq:H-1}
    \frac{\partial}{\partial t} H(t,t') = - \wh V''(h(t)) \left[ H(t,t') 
    - \d(t-t') \right] \ .
    \eeq
    The last equations must be solved with an appropriate choice of the 
    potential. We consider a sticky-sphere potential as defined in 
    Eq.~\eqref{eq:sticky_pot}, always taking $\wh v_0 =1$, 
    and will study the dynamics in the same 
    limit with $\l \to \io$ corresponding to a hard core and an infinitely
    narrow attractive region, with a constant adhesive force when $h=0$.
    
    Our goal is to compute the pair distribution function and the dynamical
    kernels based on the dynamical equations above. The first one is given 
    by~\cite{AMZ19}
    \beq\label{eq:g-1}
    g(h,t) = e^{-h} \int \de h_0 \, g_0(h_0) \, e^{h_0} 
    \moy{\d(h(t) - h)}_{h_0} \ .
    \eeq
    In our settings, this average is equivalent to the average 
    over the unitary normal variable $\x_0$. The pair distribution 
    evolution depends on the initial distribution $g_0(h_0)$; however, the 
    steady state limit must not depend on its choice, so we choose to work 
    with $g_0(h_0) = \th(h_0-1/\l)$, so that the particles are not 
    interacting at the initial time.
    
    Given these premises, the pair distribution function can be directly 
    computed in the hard-sphere limit. Indeed, when $\l \to \io$, the 
    particles are unable to overlap at $h<0$, and feel a finite attractive 
    force with strength $w_0$ when $h=0$. The trajectories can be then 
    divided into external and colliding ones. The former simply 
    follow a ballistic motion with initial velocity $\x_0$ and unitary 
    acceleration; the latter are divided in three zones: (i) a ballistic 
    motion for $t<t_1$, being $t_1$ the starting time of the collision; 
    (ii) the sticky collision, \ie $h(t)=0$ for $t_1<t<t_2$, being $t_2$ 
    the time when the particle leaves the barrier; (iii) a ballistic motion
    for $t>t_2$. Namely,
    \beq\label{eq:h-nocoll}
    h(t) = h_0 + \x_0 \, t + \frac12 t^2 \qquad \text{for external 
    trajectories,}
    \eeq
    and
    \beq\label{eq:h-coll}
    h(t) = 
    \begin{cases}
    h_0 + \x_0 \, t + \frac12 t^2 & t < t_1 = -\x_0 - \sqrt{\x^2_0-2h_0} \\
    0 & t_1 < t < t_2 = -  \x_0 + w_0 \\
    \frac12 \left[ (t+\x_0)^2 - w_0^2 \right] & t > t_2
    \end{cases}
    \qquad \text{for colliding trajectories.}
    \eeq
    At any time $t$, $h(t)$ is determined by the values of $\x_0$ and 
    $h_0$; the trajectories at contact with the barrier will contribute 
    to the delta peak in $h=0$, while the trajectories with $h>0$ will 
    give the regular part of the pair distribution function. 
    Injecting the solution above into the equation for $g(h,t)$ one has
    the time-dependent solution
    \beq\label{eq:g_sticky}
    \begin{split}
    	g(h,t) &= G(t) \, \d(h) + g_r(h,t) \ ,\\
    	G(t) &= \begin{cases}
    		\frac{t}{\sqrt{2\p}} \quad &\text{for }  t < w_0 \\
    		\frac12 + \frac{w_0}{\sqrt{2\p}} - \frac12 \erfc  \frac{t-w_0}{\sqrt 2}
    		\quad &\text{for }  t > w_0
    	\end{cases} \ , \\
    	g_r (h,t) &= \begin{cases}
    		\frac12 \left( 1+ \erf \sqrt{h} \right) + \frac{e^{-h}}{\sqrt{2\p(2h+w_0^2)}} \left[ 1 - e^{-\frac12 \left( t -  \sqrt{2h+w_0^2}\right)^2} \right]& \text{for } 0 < h<\frac{t^2 - w_0^2}2 \ , \\
    		\frac12 \left( 1+ \erf \sqrt{h} \right)& \text{for } \frac{t^2 - w_0^2}2 < h < 
    		\frac{t^2}2 \ , \\
    		\frac12 \left[1 + \erf\left(\frac{2h + t^2}{2\sqrt{2} \, t}\right)\right] & \text{for } h>\frac{t^2}2  \ , \\
    	\end{cases}
    \end{split}
    \eeq
    leading to the steady state limit for $t\to\infty$:
    \beq\label{eq:g_sticky_steady}
    g_s(h) = \Th(h) \left[ \frac12 \left( 1 + \erf \sqrt{h} \right) + 
    \frac{e^{-h}}{\sqrt{2\p(2h+w_0^2)}} \right] + \left( \frac12 + \frac{w_0}{\sqrt{2\p}} 
    \right) \d(h) \ .
    \eeq
    
    This result is equivalent to Eq.~\eqref{eq:space} in the steady state,
    and adds new information on the transient behavior of the pair 
    distribution function. In particular, the delta peak emerges continuously with time and has a singular behavior at $t=w_0$.
    When $w_0=0$, we fall back on the purely 
    repulsive hard-sphere potential studied in Ref.~\cite{ALW19}.
    \begin{figure}[!ht]
        \centering
        \includegraphics[width=0.47\textwidth]{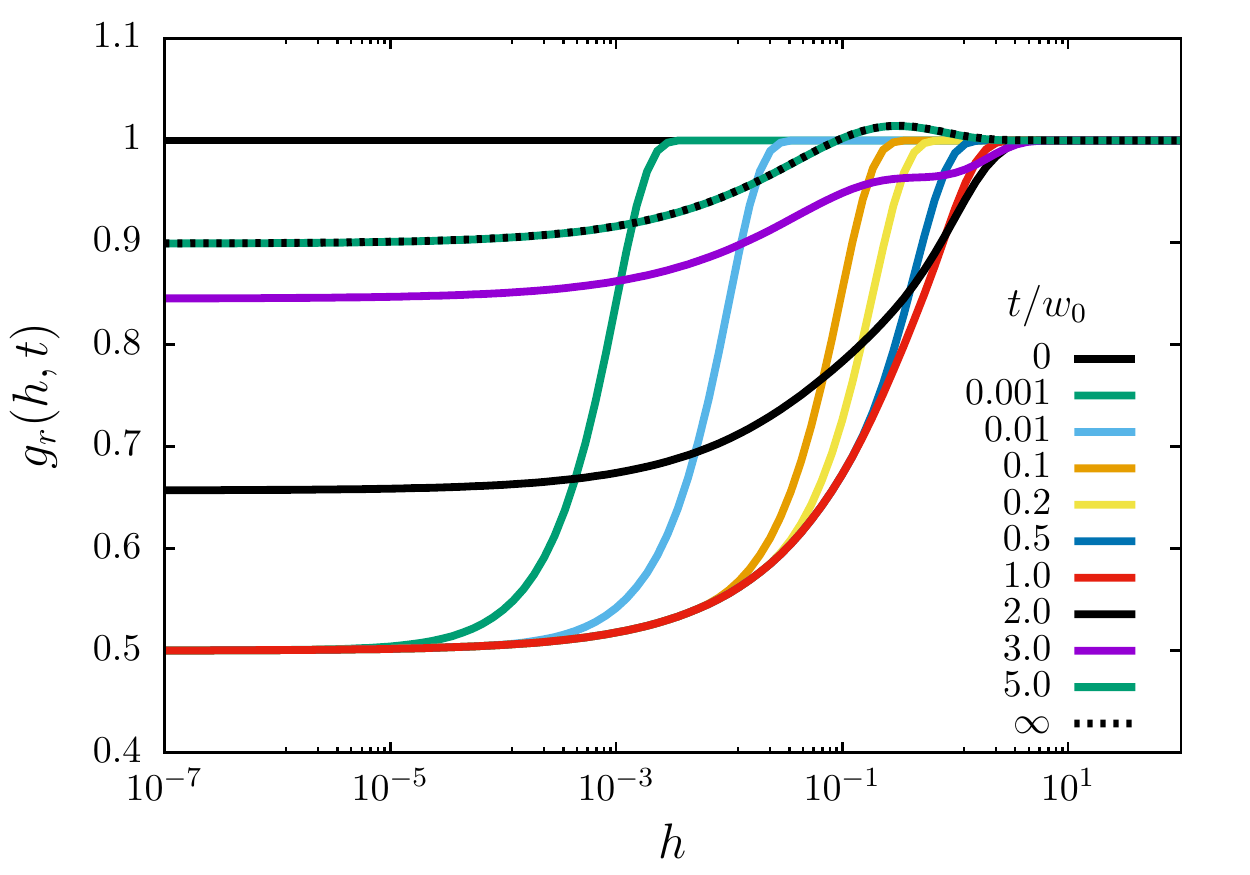}
        ~
        \includegraphics[width=0.47\textwidth]{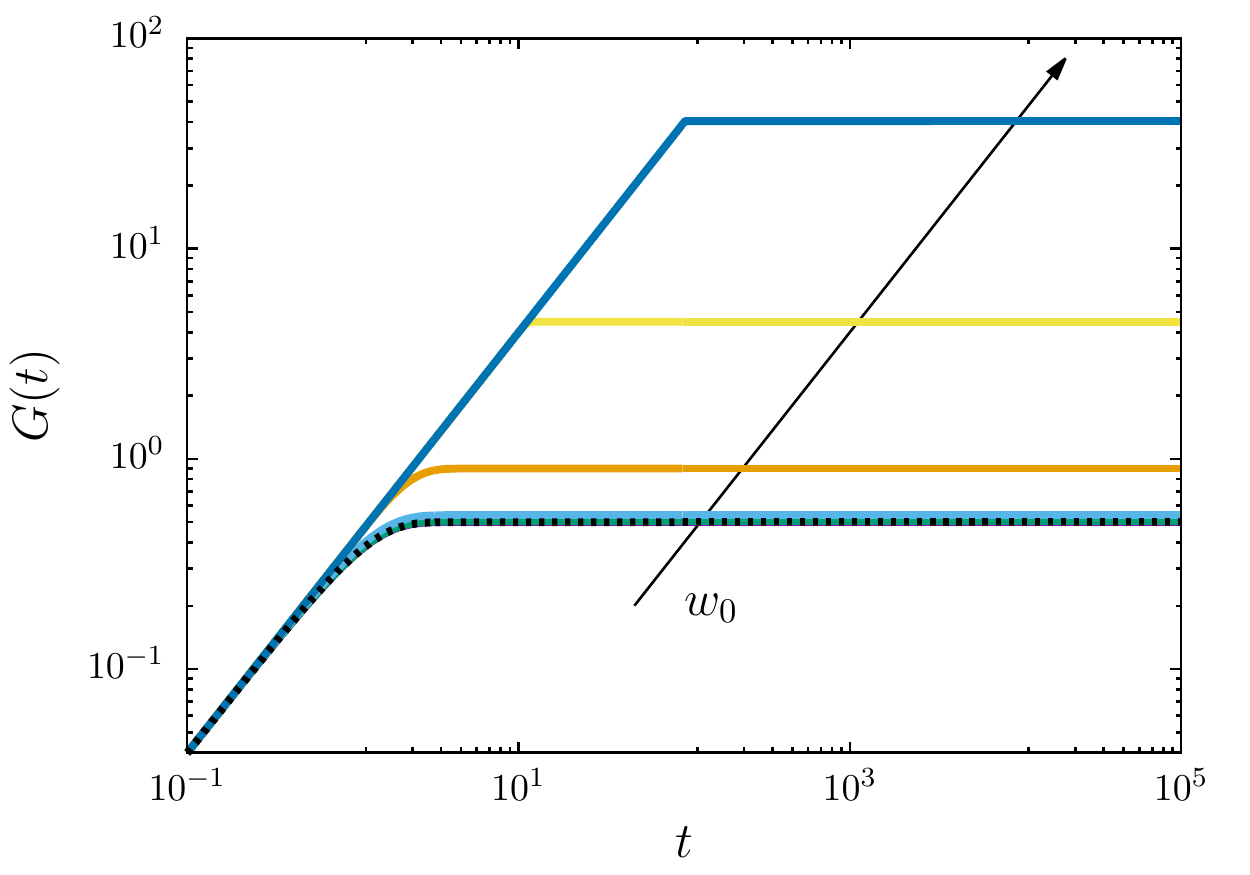}
        \caption{Left: the regular part of the pair distribution function 
        $g_r(h,t)$ vs $h$ at several times (see key), given by 
        Eq.~\eqref{eq:g_sticky} with $w_0=1$. The small-gap region 
        $h\ll 1$ is rapidly depleted by means of adhesive collisions. 
        When $t>w_0$, the self-propulsion overcomes the attractive 
        force, the particles leave the adhesive boundary and the small $h$ region becomes populated again. Right: 
        the delta peak amplitude $G(t)$ vs $t$ for $w_0=0$ (purely repulsive case, black dashed line) and $w_0=10^{-3}, 10^{-2}, 
        \ldots,10^2$ (colored lines). The linear growth at short times
        is followed by a steady state at longer times, where $G(t) \to 1/2 + w_0/\sqrt{2\p}$.
        }
        \label{fig:g}
    \end{figure}

    \subsubsection{Dynamical kernels}
    
    The computation of dynamical kernels requires the evaluation of 
    the potential and its derivatives, and thus cannot be performed in the 
    hard limit $\l \to \io$, since in that case all these terms are 
    singular. We then need to solve the equations of motion for the 
    regular potential in Eq.~\eqref{eq:sticky_pot}. Those can be solved 
    by parts for three interaction scenarios: (i) the external case $h(t)>1/\l$ for all 
    $t$, (ii) the colliding case $h(t)<0$ at some $t$ and (iii) an 
    intermediate, tangential case for which there exists $h(t)<1/\l$ but 
    $h(t)>0$ at any $t$, which means that the particles enter the mutual
    attraction region but never get to the repulsive core. This case 
    disappears in the hard potential limit, where the width of the attractive 
    region vanishes, but must be nevertheless accounted for in the course of the kernel 
    computation.
    
    The details of the computation are  reported in 
    Appendix~\ref{app:DMFT-traj}. As can be foreseen from 
    Eqs.~\eqref{eq:ker} the response kernels $\k(t)$ and $\MM_R(t,s)$ are 
    divergent in the hard-sphere limit; however, their divergences 
    compensate in that limit, as shown in 
    Appendix~\ref{app:DMFT-ker}. We also argue that in the hard-sphere
    limit the repulsive interactions give rise to a short-ranged
    memory kernel $\MM_R(t,s)$: as shown in 
    Appendix~\ref{app:DMFT-H}, the fluctuating response vanishes over a 
    time scale proportional to $\l^{-1}$ and therefore only the 
    near past of a dynamical variable contributes to the response term.
    The integrated response can be then expanded as
    \beq\label{eq:MR_exp}
    \begin{split}
    \int_0^t \de s \, \MM_R(t,s) \, f(s) &= \int_0^t \de s \, \MM_R(t,s) 
    \left[ f(t) - \dot f(t) (t-s) + \frac12 \ddot f(t) (t-s)^2 + \ldots \right] \\
    &= \c_0(t) \, f(t) - \c_1(t) \, \dot f(t) + \frac12 \c_2(t) \, 
    \ddot f(t) + \ldots \ ,
    \end{split}
    \eeq
    being $f(s)$ is a continuous function of time. The latter equation 
    is nothing but a Taylor expansion of the function $f(s)$ in the 
    integral for $s \approx t^-$, assuming that the response kernel 
    $\MM_R(t,s)$ is peaked at $s=t$ and rapidly decaying over time.
    The integrated response moments $\c_n(t)$ are defined as
    \beq\label{eq:chi-n}
    \c_n(t) \equiv \int^t_0 \de s \, \MM_R(t,s) \, (t-s)^n \ .
    \eeq
    It is shown in Appendix~\ref{app:DMFT-hw} that the moments with $n\geq2$ vanish in the hard-sphere limit. Using this property, the general motion 
    equation~\eqref{eq:y} for $y(t)$ can be approximated for $\l \gg 1$ as
    \beq
    \wh\z \dot y(t) = - \g(t) \, y(t) - \c_1(t) \,
    \dot y(t) - \wh V'(h(t)) + \X(t) \ ,
    \eeq
    where $\g(t) \equiv \k(t) - \c_0(t) $,
    and the same transformation can be applied to all the dynamical 
    equations containing the two reaction terms $\k(t)$ and $\MM_R(t,s)$.
    Their physical meaning is transparent: the first coefficient $\g(t)$ is an elastic coefficient 
    and we expect it to vanish in the long-time limit, since we are in the 
    dilute phase and the individual trajectories are not dynamically arrested near their initial 
    position. The second coefficient gives the first-order density 
    correction to the bare friction coefficient $\wh\z$, the main 
    information needed to understand how a small density affects the 
    dynamics. We also underline that this expansion does not depend 
    on the low density assumption but on the hard-sphere interactions, 
    and holds at any density.
    
    The computation of the dynamical kernels is tedious and mostly technical, 
    and is therefore deferred to Appendix~\ref{app:DMFT-ker}. It relies on 
    the computation of the two-particle process $h(t)$ and on the 
    fluctuating response $H(t,s)$, which are respectively performed in 
    Appendix~\ref{app:DMFT-traj} and~\ref{app:DMFT-H}. 
    Altogether, in the long-time limit one gets
    \beq
    \g_\io = 0 \ , \quad \c^\io_1 = \frac{\wh\f}4 \wh\z \, \left( 1 + 
    \frac{\sqrt2}{3\sqrt\p} w_0^3 \right) 
    \equiv \frac{\wh\f}{\wh\f_0 (w_0)} \, \wh\z \ .
    \eeq
    
    \subsubsection{Effective propulsion}
    
    The last result allows us to compute the effective propulsion in 
    the steady state, namely the velocity along the self-propulsion 
    direction. To do so, we write the equation for the displacement 
    of a generic particle $\d \xx(t) = \xx(t) - \xx_0$, derived through a 
    dynamical cavity method~\cite{Sz17}, before the infinite-dimensional 
    rescaling~\cite{AMZ19}. This reads
    \beq\label{eq:u}
    \begin{gathered}
    	\z \dot{\d \xx} (t) = - k(t) \d \xx(t) + \int_0^t \de s M_R(t,s) \d \xx(s) + \bff (t) + \boldsymbol{\xi}(t) \ , \\
    	\moy{f_\m(t)} = 0 \ , \quad \moy{f_\m(t) f_\n(t')} =  \d_{\m \n} \, \G_C(t-t') \ , \\
    	\moy{\x_\m(t)} = 0 \ , \quad \moy{\x_\m(t) \x_\n(t')} = \d_{\m \n} \left[ 2 \z \, T \, \d(t-t') + M_C(t,t') \right] \ ,
    \end{gathered}
    \eeq
    where we included a white thermal noise in the dynamics for the sake of generality, that we will eventually drop in the following setting $T=0$ as usual.
    We then define the dynamical observable $A(t,t')$
    \beq\label{eq:A}
    A(t,t') = \frac{\z}{v_0} \moy{ \d \xx(t) \cdot \bff (t') } \ ,
    \eeq
    measuring the total displacement at time $t$ along the 
    direction of the active force at time $t'$. This quantity leads 
    to the definition of the effective propulsion $v(\wh\f)$ as
    \beq\label{eq:v_rho_def}
    v(\wh\f) = \left. \frac{\partial}{\partial t} A(t,t') \right\vert_{t=t'} = 
    \frac1{v_0} \moy{\z \dot{\d \xx} (t) \cdot \bff (t)}  \ ,
    \eeq
    which is a time-translationally invariant observable in the steady state.
    So, at zero density, the free-particle is moving at the 
    bare self-propulsion speed $v_0$, and we expect $v(\wh\f)$ to 
    decrease monotonically with the density.
    
    The quantity $A(t,t')$ follows the dynamical equation
    \beq\label{eq:dA}
    \partial_t A(t,t') = - k(t) A(t,t') + \int_0^t \de s \, 
    M_R(t,s) A(s,t') + \frac{d \z}{v_0} \G_C(t-t') \ ,
    \eeq
    having exploited the independence between the active noise and the 
    fluctuations of the inter-particle interactions, \ie $\moy{\bxi(t) \cdot \bff (t')}=0$.
        
    When $t,t' \to \io$, we obtain the steady-state dynamical equation
    \beq\label{eq:dA_steady}
    \z \partial_t A(t,t') = - g(t) A(t,t') - c_1(t) \partial_t A(t,t') 
    + \frac{d \z}{v_0} \G_C(t-t') \ ,
    \eeq
    where $g(t) = k(t) - \int^t_0 \de s \, M_R(t,s) = (2d^2/\ell^2) \g(t) 
    \to 0$ and $c_1(t) = \int^t_0 \de s \, M_R(t,s) \, (t-s) = (2d^2/\ell^2) 
    \c_1(t) \to c^\io_1$ when $t\to\io$. The last results hold for 
    $\G_C(t-t') = v_0^2/d$, and the total friction coefficient 
    reads $\z + c_1^\io = \z (1+ \wh\f/\wh\f_0)$. Their derivation 
    is presented in Sec.~\ref{app:DMFT-ker}, and is given by 
    the computation of the first two integrated response moments in 
    the limit of dilute hard spheres.
    So, Eq.~\eqref{eq:v_rho_def} gives us
    \beq\label{eq:v-phi}
    v(\wh\f) = \frac{v_0}{1+\wh\f/\wh\f_0 (w_0)} \ .
    \eeq
    This is the fundamental result of this calculation. We show then that, 
    to the first order in $ \wh\f$, the effective propulsion in a 
    dilute media is damped by a factor $1 + \wh\f/\wh\f_0 (w_0)$, 
    accounting for the slowing down of particles' velocity caused by 
    interactions. Considering the dilute limit approximation
    $\wh\f \ll 1$, its first-order expansion in $\wh\f$ coincides with 
    the result obtained from Fokker-Planck equation in Eq.~\eqref{eq:veff}.
    
    It is worth noticing that the last result can be generalized to relate the 
    friction coefficient and the effective self-propulsion with the steady-state
    pair distribution function. Indeed from Eq.~\eqref{eq:chi-n} one also has
    \beq
    \c^\io_1 = \frac{\wh\f}2 \int^{+\io}_{-\io} \de h \, e^h g_s (h) \KK^\io(h) \ , \quad \KK(h,t) = \int_0^t \de s \, (t-s) \moy{ \wh V''(h(t)) H(t,s)}_{h}  \ ,
    \eeq
    being $\KK^\io(h) = \lim_{t\to\io} \KK(h,t)$.
    
    \section{Transient behavior of Hard Spheres}
    \label{sec:transient}
    
    When $w_0=0$, we recover the purely repulsive hard-sphere interaction
    potential, namely
    \beq\label{eq:vHS}
    \wh V_{\rm HS} (h) = 
    \begin{cases}
    \io & h<0 \\
    0 & h>0
    \end{cases}
    \ .
    \eeq
    All calculations above are valid for the case $w_0=0$. Furthermore,
    in this case one can also compute the transient dynamics of the 
    dynamical kernels defined in Eqs.~\eqref{eq:ker}, which were 
    analytically unattainable in the general sticky spheres case. With 
    the same procedure as in the previous section, we approximate the hard-sphere 
    potential with a soft-sphere one, namely $\wh V(h) = \dfrac{\ee}2 h^2 
    \th(-h)$. The hard-sphere potential is recovered in the $\ee\to\io$ limit. 
    This soft-sphere potential is equivalent to the sticky-sphere one 
    defined in Eq.~\eqref{eq:sticky_pot}, in the limit $w_0 \to 0$ and $\l \to \io$ 
    keeping $\l w_0 = \ee$ fixed, and it is the same interaction potential 
    already analyzed in the solution of equilibrium dynamics presented 
    in~\cite{MSZ20}.
    
    This choice makes the analytical computation of the dynamical kernels
    much easier; indeed, one can follow the same scheme described for 
    sticky spheres to access the pair distribution function $g(h,t)$ and 
    the dynamical kernels $\k(t)$, $\MM_R(t,s)$ and $\MM_C(t,s)$. The 
    time evolution of the pair distribution function is given by Eq.~\eqref{eq:g_sticky}, 
    setting $w_0=0$. For the dynamical kernels, 
    we can avoid the limit $t\to\io$ in their calculation; we get then 
    the first integrated response moments, finally leading to
    \beq\label{eq:HS-ker}
    \begin{split}
        \g(t) &= \frac{\wh\f}2  \wh v_0 \, \left[ \frac{e^{-t^2/2}}{\sqrt{2\p}} - 
        \frac{t}2 \, \erfc \left( \frac{t}{\sqrt2} \right) \right] \ , \\
        \c_1(t) &= \frac{\wh\f}4  \wh\z \, \erf \left( \frac{t}{\sqrt2} \right) \ , 
        \\
        \MM_C(t,s) &= \frac{\wh\f}{4}  \wh v_0^2 \, \left[ \erfc \left( \frac{\vert t 
        - s \vert}{\sqrt{2}} \right) + \erfc \left( \frac{t}{\sqrt{2}} 
        \right) (1+t s) - \sqrt{\frac{2}{\p}} s \, e^{-t^2/2} \right] \ ,
    \end{split}
    \eeq
    always expressing the time $t$ in units of the natural time scale 
    $\t_0=\wh\z/\wh v_0$. The steady-state limit is the same as that described for the sticky-sphere 
    case, \ie $\g_\io = 0$ and $\c^\io_1 = \dfrac{\wh\f}4 \, \wh\z \equiv 
    \dfrac{\wh\f}{\wh\f_0} \, \wh\z$. Furthermore,
    we can characterize the noise correlation in the long-time limit, 
    where the noise only depends on the time difference $\t = t-s$, 
    namely
    \beq\label{eq:HS-MC}
    \MM^\io_C(\t) = \frac{\wh\f}4 \wh v_0^2 \, \erfc \frac{\vert \t 
    \vert}{\sqrt2} \ .
    \eeq
    
    The above result allows us to derive the behavior of the MSD in the 
    long-time limit; indeed, with the kernels computed in 
    Eq.~\eqref{eq:HS-ker}, the correlation-response equations now read
    \beq\label{eq:CR-1-HS}
    \begin{split}
    \wh\z \frac{\partial}{\partial t} \cR(t,t') &= \frac{\d(t-t')}2 - 
    \g(t) \cR(t,t') - \c_1(t) \frac{\partial}{\partial t} \cR(t,t') \ , \\
    \wh\z \frac{\partial}{\partial t} \CC(t,t') &= - \g(t) \CC(t,t') 
    - \c_1(t) \frac{\partial}{\partial t} \CC(t,t') + \int^{t'}_0 \de s \, 
    \left[ \GG_C(t-s) + \MM_C(t,s) \right] \cR(t',s) \ .
    \end{split}
    \eeq
    The first equation can be explicitly solved in the steady-state limit, 
    giving
    \beq\label{eq:R-HS-ss}
    \cR^\io (\t) = \frac1{2\wh\z (1+\wh\f/\wh\f_0)} \th(\t) \ .
    \eeq
    With this result, from Eq.~\eqref{eq:MSD} one can derive an equation 
    for the MSD in the steady-state limit $t,t'\gg 1$, as a function 
    of the dimensionless time difference $\t = t-t'$, \ie
    \beq\label{eq:MSD-HS-ss}
    \dot \D^\io (\t) = \frac1{(1+\wh\f/\wh\f_0)^2} 
    \left\{ \t + \frac{\wh\f}{\wh\f_0} \left[ \t \, \erfc 
    \frac{\t}{\sqrt2} + \sqrt{\frac2{\p}} \left(1 - e^{-\t^2/2}\right) \right] \right\} \ ,
    \eeq
    and this equation can be easily integrated, yielding
    \beq\label{eq:Delta}
    \D^\io(\t) = \frac1{(1+\wh\f/\wh\f_0)^2} 
    \left\{ \frac{\t^2}2 + \frac{\wh\f}{\wh\f_0} \left[ \frac{\t}{\sqrt{2\p}} \left( 2 - 
    e^{-\t^2/2} \right) - \frac12 \erf 
    \frac{\t}{\sqrt2} + \frac{\t^2}2 \erfc \frac{\t}{\sqrt2} \right] 
    \right\} \ .
    \eeq
    \begin{figure}[!ht] 
        \centering
        \includegraphics[width=0.47\textwidth]{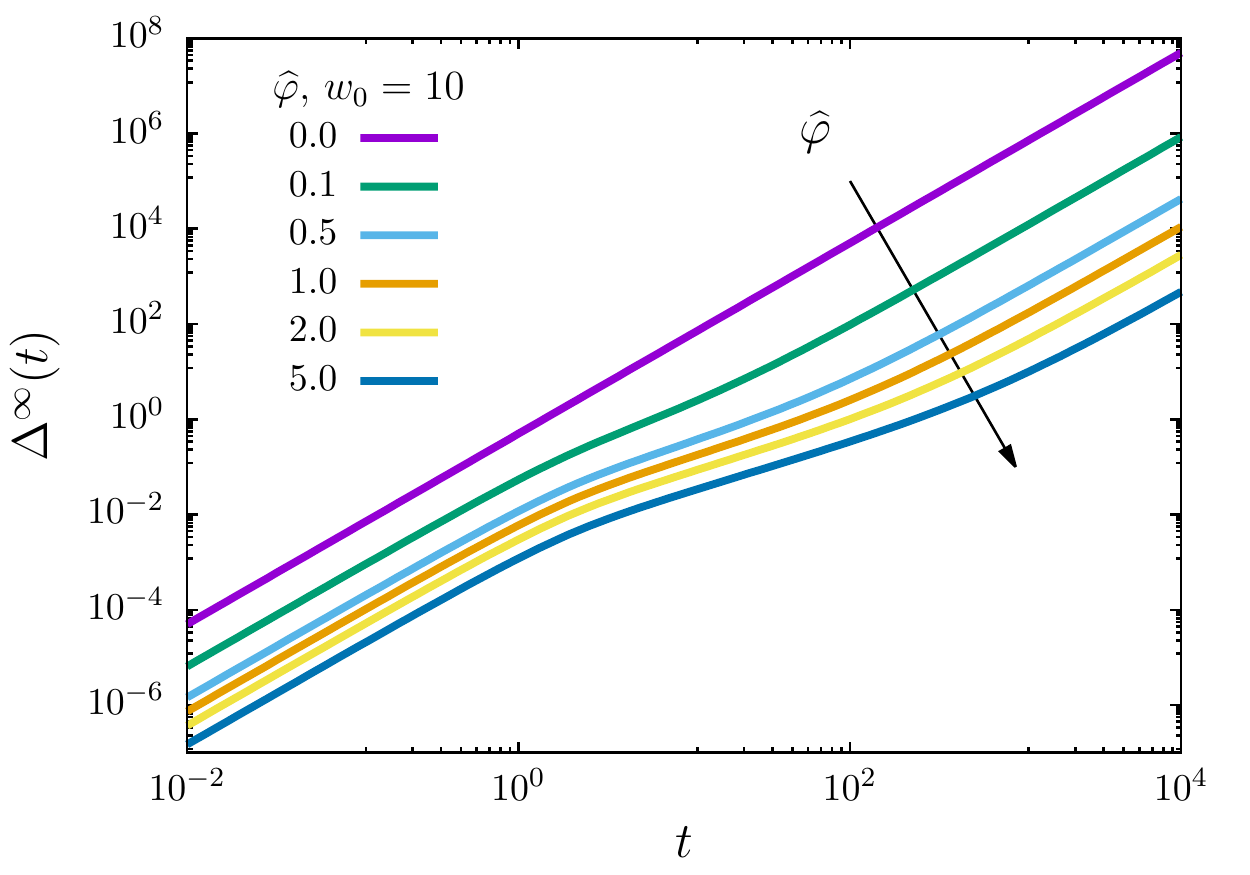}
        ~
        \includegraphics[width=0.47\textwidth]{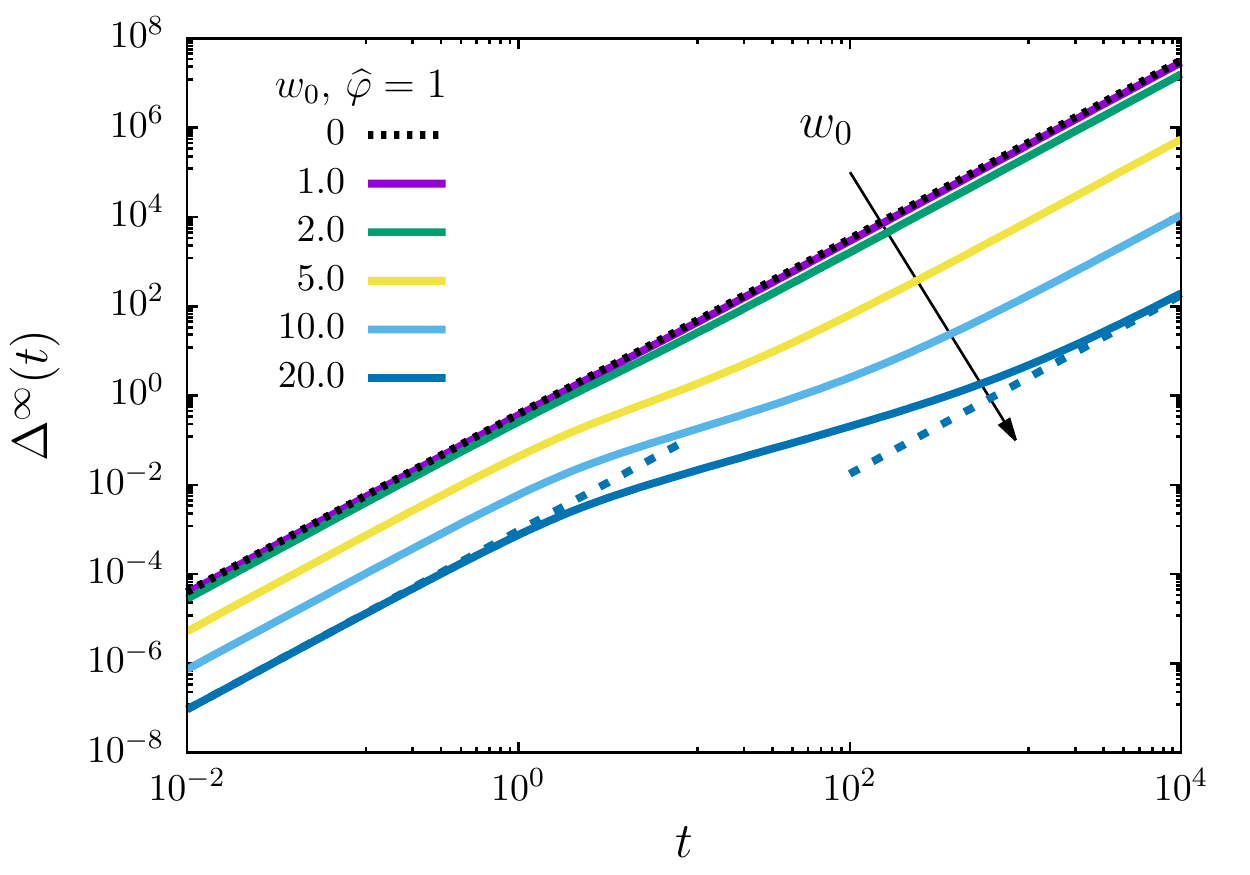}
        \caption{Steady-state mean squared displacement $\D^\io(t)$ vs $t$ from 
        Eq.~\eqref{eq:Delta} for several 
        values of rescaled density $\wh\f$ (left, $w_0=10$) and attractive 
        force $w_0$ (right, $\wh\f=1$). The MSD is ballistic at short and long times, 
        but the increase in density or in adhesion induces 
        a slowdown at intermediate times, respectively given by the 
        many-body interactions or the duration of an adhesive collision. 
        We compare 
        the short and long time behavior from Eq.~\eqref{eq:Delta-asym} 
        for the case $w_0=20$ in the right panel (dashed blue lines).}
        \label{fig:Delta}
    \end{figure}
    The solution above shows that the dynamics is ballistic at short and 
    long times, with a slowdown at intermediate times given by the 
    presence of interactions. With the effective propulsion definition
    computed in Eq.~\eqref{eq:v-phi}, it is clear that
    \beq\label{eq:Delta-asym}
    \D^\io(\t) \sim \frac{v(\wh\f)^2}{2 \wh\z^2}  \t^2 \times 
    \begin{cases}
    1 + \wh\f/\wh\f_0 & \t \ll \t_0 \\
    1 & \t \gg \t_0
    \end{cases}
    \eeq
    with $\t_0=\wh\z/\wh v_0$.
    The last result can be interpreted in the following way: at any time, 
    the interplay between self-propulsion and pairwise interactions
    yields an effective self-propulsion $v(\wh\f)$ for the single particle, 
    hence the common prefactor in the rhs of Eq.~\eqref{eq:Delta-asym}.
    At short times, however, the fluctuations of pairwise interactions $\bxi(t)$
    are correlated and therefore contribute to the MSD with an additional 
    term as shown above. At longer times, conversely, the fluctuations always
    decorrelate in the dilute phase and the dominant contribution to the MSD 
    is given by the effective self-propulsion.
    
    \section{Conclusions}
    
    In this paper, we implemented two alternative approaches to 
    investigate the dynamics of active particles embedded in 
    a large-dimensional space. The results from kinetic theory, previously 
    presented in~\cite{ALW19}, have been extended to the case of a 
    non-monotonous potential featuring an attractive component, and 
    compared with the results of dynamical mean-field theory in the 
    low-density limit and at large persistence time. The two methods have 
    been proven to be consistent in the steady-state limit within this frameword; 
    indeed, the calculation of the pair distribution function has given the same 
    result in both cases, and we added a more detailed description in 
    term of the rescaled inter-particle gap $h(t)$, which is the 
    physically interesting variable in the high-dimensional limit. 
    Finally, we have explicitly computed the effect of an infinitely 
    short-ranged attractive potential on the amplitude of the adhesive 
    delta peak in $g(h)$.
    
    Furthermore, we also computed the effective propulsion 
    $v(\wh\f)$ of the active particles, namely the effective velocity at 
    which particles are propelled once the effect of the interactions have 
    been considered. Again, the two approaches converged to the same 
    results, taking into account that this result is obtained in the 
    dilute 
    limit. The extrapolation of the effective self-propulsion to higher 
    densities can lead to interpret $\wh\f_0(w_0)$ as the crowding 
    density, \ie the value of the density at which the active particles 
    get stuck in an arrested phase with 
    no room to move any longer~\cite{ALW19}. We remark that for purely repulsive hard 
    spheres one has $\wh\f_0 = 4$, and this value decreases 
    when adhesion is present. On the other hand, it is 
    known~\cite{parisi2020theory} that hard spheres in equilibrium undergo a 
    dynamical glass transition at $\wh\f = \wh\f_d \simeq 4.8067$, where the 
    mean square displacement converges to a finite but positive limit; 
    conversely, both the effective propulsion and the 
    mean square displacement vanish in the crowded phase. We stress that 
    the result for the crowding density is however extrapolated from the dilute phase, and that it must 
    be taken as a first step towards a solution in the dense phase, where 
    the crowding and glass transitions can be compared properly.
    
    The last part of our analysis has been dedicated to the transient behavior 
    of hard spheres by means of dynamical mean-field theory; it has been 
    shown how, starting from an equilibrium configuration, the system relaxes 
    towards a stationary state. This relaxation is described by the transient 
    part of the dynamical coefficients, and in this limit we computed the 
    MSD in the steady state, elucidating how the interplay between active 
    self-propulsion and interactions affects its short-time behavior, while 
    the infinitely persistent self-propulsion dominates at long times.
    
    These results constitute a starting point for a more complete analysis of active 
    systems in high dimensions. The next step along this line of research is its 
    extension to higher densities and finite persistence times. This task being 
    severely hard to accomplish via analytical tools, a numerical solution of DMFT 
    equations must be found, in line with previous results~\cite{RBBC19,MSZ20}.
    However, if the self-propulsion is too strong or too persistent, the 
    trajectories drift away and the solution relies on the 
    statistics of exponentially rare events. The development of importance-based 
    algorithms is then required and would give an important edge in the solution 
    of the problem at any density.
    
    Another approach that may be tackled in the future concerns the limit of 
    small persistence time; in that case, often studied in active matter 
    systems~\cite{FNCTVvW16prl,FS20pre} the dynamics can 
    be perturbatively studied starting from the equilibrium solution.
    Its analysis would lead to understand how a small amount of activity affects 
    the dynamics, \ie the behavior of dynamical kernels, the interplay between 
    the dynamical transition and the crowding transition, and the effects 
    on fluctuation-dissipation relations.
    
\acknowledgments

We thank E. Agoritsas and L. Berthier for interesting discussions related to this work.
This project has received funding from the European Research Council (ERC) under the European Union's Horizon 2020 research and innovation programme (grant agreement n. 723955 - GlassUniversality). TAP and FvW have been supported by the ANR THEMA funding.

\section*{Data Availability}    
    
    Data sharing is not applicable to this article as no new data were created or analyzed in this study.
    
    \bibliographystyle{ieeetr}
    \bibliography{active_dinf.bib}
    
    \clearpage
    
    \appendix
    \section{Solving the two-body Fokker-Planck equation}
    \label{app:FP}
    
\begin{center}
    \begin{figure}[!ht]
    \begin{overpic}[totalheight=6cm]{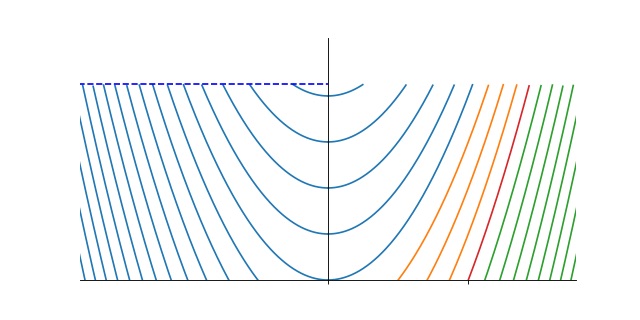}
     \put (72,3) {$w_0$}
     \put (50.65,3) {$0$}
     \put (52,36.5) {$L$}
    \end{overpic}
    \caption{Characteristics in the ($w, h$) plane.}
    \label{fig:charac}
    \end{figure}
    \end{center}

    In this Appendix, we solve Eq.~\eqref{eq:FP_ultraballistic} using the method of characteristics for the sticky-sphere potential. We start by establishing Eq.~\eqref{eq:bulkFP} and Eq.~\eqref{eq:boundaryFP} of the main text which describe the $\lambda \to \infty$ limit of the stationary distribution. For $h > 1/\lambda$, we obtain first
    \begin{align}
        w \, \partial_h P + \partial_w P = 0 \, .
    \end{align}
    In the limit $\lambda \to \infty$, we thus recover Eq.~\eqref{eq:bulkFP} of the main text. Next, for $h < 1/\lambda$ and for any function $j(h)$ independent of $\lambda$ we define 
    \begin{align}
        \Gamma^\lambda_j(w) = \int_{-\infty}^{1/\lambda} \dd h \, e^{h} P(h, w) j(h) \, ,
    \end{align}
    so that Eq.~\eqref{eq:FP_ultraballistic} yields
    \begin{equation}\label{eq:integration}
        - \wh{v}_0 \partial_w \Gamma^\lambda_j(w) + \wh{v}_0 w \Gamma^\lambda_j(w) - \wh{v}_0 w \left( P\left(1/\lambda, w\right) j\left(1/\lambda\right) e^{1/\lambda}\right) + \wh{v}_0 w \Gamma^\lambda_{j'}(w) - \int_{-\infty}^{1/\lambda} \dd h \, e^h \frac{\wh{V}'(h)}{\ell}j'(h) P(h,w) = 0 \,.
    \end{equation}
    In the limit $\lambda \to \infty$, the stationary distribution function decays to $0$ as $h < 0$ over scales $O(1/\lambda)$ and we have
    \begin{align}
        \Gamma^\lambda_j(w) \xrightarrow[\lambda \to \infty]{} j(0) \lim_{\lambda \to\infty} \int_{-\infty}^{1} \frac{\dd h}{\lambda} \,  e^{h/\lambda} \, P\left(\frac{h}{\lambda}, \, w\right) = f(0) \, \Gamma(w) \, ,
    \end{align}
    provided the previous limit exists. This justifies the functional form in Eq.~\eqref{eq:stat_prob} of the main text. Hence, on one hand, for a function $j$ defined such that $j'(0) = 0$, Eq.~\eqref{eq:integration} yields Eq.~\eqref{eq:boundaryFP}
    \begin{equation}
        \Gamma'(w) - w \Gamma(w) = - w \lim_{\lambda \to \infty}P\left(1/\lambda, w\right) \, .
    \end{equation}
    On the other hand, for a function $j$ such that $j(0) = 0$, Eq.~\eqref{eq:integration} yields the integrated version of Eq.~\eqref{eq:regular}
    \begin{equation} \label{eq:regular_integrated}
        \lim_{\lambda \to \infty} \int_{-\infty}^{1/\lambda} \dd h \, e^{h} j'(h) \frac{\wh{V}'(h)}{l}P(h,w) = \wh{v}_0 w j'(0) \Gamma(w)\ ,
    \end{equation}
    which gives the limit of the product $\wh{V}'(h) P(h,w)$ as $\lambda \to \infty$. Eventually, since $\wh{V}'(h) < \wh{v}_0 w_0$, we obtain from Eq.~\eqref{eq:regular_integrated}
    \begin{equation}
        \Gamma(w) \left(w - w_0\right) \leq 0 \, ,
    \end{equation}
    which, given the positivity of $\Gamma(w)$, yields
    \begin{equation}
        \Gamma(w > w_0) = 0 \, .
    \end{equation}
    We are now in position to solve Eq.~\eqref{eq:bulkFP} and Eq.~\eqref{eq:boundaryFP}. In Sec.~\ref{sec:DMFT-g}, the same stationary distribution will be derived in an alternative way directly from the equations of motion. For $h > 0$, Eq.~\eqref{eq:bulkFP} tells us that $P^b(h,w)$ is constant along the characteristics $2h - w^2 = \text{cst}$ that correspond to deterministic trajectories. These characteristic lines are depicted in Fig.~\ref{fig:charac}. We solve the equations with the boundary condition 
    \begin{equation} \label{eq:bc}
       P^b(L, x < 0) = 1 \, ,
    \end{equation}
    where $L$ is some large length scale introduced to treat the boundary conditions that will eventually be sent to infinity. In the relative-particle-around-a-spherical-obstacle picture this corresponds to a homogeneous reservoir of incoming particles at $h = L$. As $L$ is sent to infinity this expresses the isotropy of the stationary distribution at large distances. 
    The blue domain in Fig.~\ref{fig:charac}, \textit{i.e.} $\{w < 0, h\} \cup \{ w > 0, 2h - w^2 > 0\}$, is made of characteristics that intersect the boundary half line $\{ x < 0, h = L\}$. The quantity $P^b$ is thus constant and equal to one in this domain. On the contrary, $P^b$ vanishes in the orange ($\{w>0, 0 > 2h - w^2 > - w_0^2\}$) and green ($\{w>0, 2h - w^2 < - w_0^2\}$) ones. Indeed, we have first $\Gamma(w > w_0) = 0$ so that Eq.~\eqref{eq:boundaryFP} implies $P^b(0, w > w_0) = 0$ and the vanishing of $P^b$ in the green domain. Then, we notice that the orange domain corresponds to trajectories in which the two particles escape from a collision event with $0 < w < w_0$. However, given the shape of the potential in Eq.~\eqref{eq:sticky_pot}, this never happens. Eventually, the red line $2h - w^2 = - w_0^2$ in Fig.~\ref{fig:charac} plays a special role. Indeed, all trajectories leading to a collision event between the two particles collapse onto this line as they separate afterwards. We thus look for a solution of Eq.~\eqref{eq:bulkFP} of the form
    \begin{align}
        P^b(h, w) = P_0(h, w) + f(h)\,\delta\left(h - \frac{w^2}{2} + \frac{w_0^2}{2}\right)\Theta(w) \, ,
    \end{align}
    with $P_0(h,w)$ a piece-wise continuous function whose form was derived above. Equation~\eqref{eq:bulkFP} then yields
    \begin{align}
        f'(h) = 0 \Rightarrow f(h) = f(0) \, .
    \end{align}
    The constant $f(0)$ is then found by integrating Eq.~\eqref{eq:boundaryFP} between $w_0^-$ and $w_0^+$. This yields
    \begin{align}
        f(0) = \Gamma(w_0^-) \, .
    \end{align}
    We are now in position to solve Eq.~\eqref{eq:boundaryFP}. For $w < 0$, $P^b(0, w) = 1$ and we obtain
    \begin{equation}
        \Gamma'(w) - w \Gamma(w) = - w \, ,
    \end{equation}
    so that
    \begin{equation}
        \Gamma(w) = A e^{w^2/2} + 1 \, ,
    \end{equation}
    with $A$ an integration constant that is set to $0$ to ensure the integrability of $\Gamma(w)$ against $e^{-w^2/2}$. For $0 < w < w_0$, we have $P^b(0, w) = 0$ and thus
    \begin{align}
        \Gamma(w) = e^{w^2/2} \, ,
    \end{align}
    where the integration constant was chosen to ensure continuity at $w = 0$. We have therefore derived Eq.\eqref{eq:bulk}
    \begin{align}
        P_b(h,w) = \Theta(h)\left[1 - \Theta(w)\Theta\left(\frac{w^2}{2}-h\right) + \Theta(w)e^{\frac{w_0^2}{2}}\delta\left(h - \frac{w^2}{2} + \frac{w_0^2}{2}\right)\right]
    \end{align}
    and Eq.~\eqref{eq:surface}
    \begin{align}
        \Gamma(w) = \Theta(-w) + \Theta(w)\Theta(w_0-w)e^{\frac{w^2}{2}}
    \end{align}
    of the main text.

    \section{Dynamical Mean-Field with a sticky potential}
    
    \subsection{Trajectories}
    \label{app:DMFT-traj}
    
    Here we solve the equation of motion of the rescaled gap $h(t)$ as 
    defined in Eq.~\eqref{eq:h-1}, for a sticky potential such as the one 
    defined in Eq.~\eqref{eq:sticky_pot}.
    
The trajectories start as
\beq\label{eq:h01}
h_{01} (t) = h_0 + \x_0 t + t^2/2 \ ,
\eeq
so, the attractive region is reached when $h(t) = 1/\l$ at time 
\beq\label{eq:t1}
t_1 = - \x_0 - \sqrt{\x^2_0 - 2(h_0-1/\l)} \ .
\eeq
This happens for $(\x_0,h_0)$ such that
\beq\label{eq:region1}
\x_0 < 0 \quad \wedge \quad h_0 < \frac1{\l} + \frac{\x^2_0}2 \ .
\eeq
For all other values of $(\x_0,h_0)$, the trajectory always stays  in the noninteracting region.\\
We therefore define the new variable $\a = \sqrt{\x^2_0 - 2(h_0-1/\l)} > 0$, so that
\beq\label{eq:t1_a}
t_1 = - \x_0 - \a \ .
\eeq
The condition $h_0 > 1/\l$ implies $\a < |\x_0|$, and since $\x_0 < 0$ for the trajectories 
of our interest we have
\beq\label{eq:alpha}
0 < \a < -\x_0 \quad \wedge \quad \x_0 < 0 \qquad \Leftrightarrow \qquad \a > 0 \quad \wedge 
\quad \x_0 < -\a \ .
\eeq

%
%
Finally, the weight in the integrals reduces to
\beq\label{eq:exp_xi_h0}
\DD \x_0 \, \de h_0 \, e^{h_0} = \frac1{\sqrt{2\p}} \de \x_0 e^{-\x^2_0/2} \, \a \, \de \a \, 
e^{1/\l +\x^2_0/2 - \a^2/2} = \frac{e^{1/\l}}{\sqrt{2\p}} \, \de \x_0 \, \a \, \de \a \, e^{-\a^2/2} \ .
\eeq

\subsubsection{Tangential trajectories}

Assuming we have entered the attractive region, we have the Cauchy problem
\beq\label{eq:hdiff12}
\left\lbrace
\begin{aligned}
	\dot h(t) &= \l w_0 h(t) - w_0 + \x_0 + t \\
	h(t_1) &= \frac1{\l}
\end{aligned}
\right. \ ,
\eeq
which is valid for $0<h(t)<1/\l$. The analytical solution is
\beq\label{eq:h12}
h_1 (t) = \frac1{(\l w_0)^2} \left[ -1 + \l w_0^2 + \l w_0 \a  -\l w_0 (t-t_1) + e^{\l w_0 (t-t_1)} 
\left( 1 - \l w_0 \a  \right) \right] \ .
\eeq
For later convenience, we define $z = \l w_0 \a - 1$ and $x = t-t_1$. Thus, we can rewrite
\beq\label{eq:h12_z}
h_1 (x=t-t_1) = \frac1{\l w_0} \left[ w_0 - x - \frac{z}{\l w_0} \left( e^{\l w_0 x} - 1 \right) \right] .
\eeq
There are now two possibilities:  
\begin{enumerate}
	\item the trajectory is tangential, $h_{\min}(t) > 0$ and therefore it crosses the attractive 
	region and leaves it at a given time $t_5$;
	\item the trajectory is colliding, therefore there is a positive $x=t-t_1$ at which $h(t)=0$.
\end{enumerate}
We need to solve the equation $h_1(t)=0$. Its solution leads to
\beq\label{eq:t2}
t-t_1 = \frac1{\l w_0} \left[ \l w_0^2 + z - W\left( z e^{\l w_0^2+z} \right) \right] \ ,
\eeq
where $W(x)$ is the Lambert function. The Lambert function has one branch for $x>0$ and 
two branches for $-e^{-1}<x<0$. Therefore, $x$ exists if 
\beq
ze^{\l w_0^2+z} > - e^{-1} \quad \Rightarrow \quad z e^z > - e^{-1-\l w_0^2} \quad \Rightarrow 
\quad  z < z_1 = W_{-1} \left( - e^{-1-\l w_0^2} \right) \vee z>z_2 = W_0 \left( - e^{-1-\l w_0^2} 
\right) 
\eeq
Given $z = \l w_0 \a -1 > -1$ and $z_1<-1$, the \textit{colliding condition} reduces to $z>z_2$.
We also note that $x = t-t_1 >0$ for every $z$ satisfying this condition. Indeed, if 
$z<0$ there are two possible values of $x$, corresponding to the fact that the coefficient of 
the exponential in Eq.~\eqref{eq:h12_z} is positive and therefore the \textit{virtual} trajectory 
would cross the barrier twice and then diverge to $+\io$; in this case, the primary branch 
of the Lambert function corresponds to the first intersection and the secondary branch to the 
second one.\\
On the other hand, if $z>0$ there is only one intersection with the barrier because the virtual 
trajectory diverges to $-\io$, corresponding to the unique branch of $W(x)$ for $x>0$.

This result further divides  the $(\x_0,z)$ plane into the following cases
\beq
\begin{split}
	-1 < z < z_2 \quad &\Rightarrow \quad \text{tangential trajectory} \\
	z > z_2 \quad &\Rightarrow \quad \text{colliding trajectory}
\end{split}
\eeq
Having found the values of $z$ for which the trajectory is tangential, we can now compute 
the exit time $t_5$ ($t_i$ with $i=2,3,4$ will be reserved for colliding trajectories): we need 
indeed to solve the equation
\beq\label{eq:dt15}
h_1(t) = \frac1{\l}  \quad \Rightarrow \quad \d t_{15}(z) = t_5 - t_1 = \frac1{\l w_0} \left[ z - W_{-1} \left ( z e^z \right) \right] \ .
\eeq
The two branches of $W(x)$ give two solutions: since $-1 < z < 0$, we have that $W_0 (z e^z) 
= z$ and the solution above gives the trivial result $t_5 = t_1$; the second branch gives 
$W_{-1}(z e^z) < z$ and therefore a positive result for $\d t_{15} (z)$.

So we have the trajectory from $t_1$ (entrance time) to $t_5$ (exit time), where the particle
crosses the attractive region and contributes to the kernels.

\subsubsection{Colliding trajectories}

We now compute the colliding trajectories, which require $\x_0 < -\a = - (1+z)/(\l w_0)$ and 
$z > z_2$.

\paragraph{Attractive region 1: zone 12}

The motion in the attractive region towards the barrier has been already computed in 
Eq.~\eqref{eq:h12_z}. We recall the trajectory from Eq.~\eqref{eq:h12_z} and the colliding time 
$t_2$ from Eq.~\eqref{eq:t2}, \ie
\beq
h_1 (x=t-t_1) = \frac1{\l w_0} \left[ w_0 - x - \frac{z}{\l w_0} \left( e^{\l w_0 x} - 1 \right) \right] .
\eeq
\beq
\d t_{12}(z) = t_2 - t_1 = \frac1{\l w_0} \left[ \l w_0^2 + z - W_0 \left( z e^{\l w_0^2+z} \right) \right]\ . 
\eeq
\vspace{5mm}

\paragraph{Repulsive region: zone 23}

The motion in the repulsive region needs the solution of the Cauchy problem
\beq\label{eq:hdiff23}
\left\lbrace
\begin{aligned}
	\dot h(t) &= - \l w_0 h(t) - w_0 + \x_0 + t \\
	h(t_2) &= 0
\end{aligned}
\right.
\eeq
Its solution reads 
\beq\label{eq:h23}
h_2 (x = t-t_2) = \frac1{\l w_0} \left[ x - \frac{w(z)}{\l w_0} \left( 1 - e^{-\l w_0 x} \right)  \right] \ ,
\eeq
being $w(z) = 2 + W_0 \left( z e^{\l f^2_0 + z} \right)$.

Exit time:
\beq\label{eq:dt23}
\d t_{23}(z) = t_3 - t_2 = \frac1{\l w_0} \left[ w(z) + W_0 \left(-w(z) e^{-w(z)} \right) \right] \ .
\eeq
Since $ - w(z) < -1$, then $W_{-1}(-w e^{-w}) = -w$ so the secondary branch gives the trivial 
solution $t_3 = t_2$; therefore we choose the primary branch $W_0$ into Eq.~\eqref{eq:dt23}.

\paragraph{Attractive region 2: zone 34}

For $t>t_3$, the particle enters back the attractive region, \ie
\beq\label{eq:hdiff34}
\left\lbrace
\begin{aligned}
	\dot h(t) &= \l w_0 h(t) - w_0 + \x_0 + t \\
	h(t_3) &= 0
\end{aligned}
\right.
\ ,
\eeq
yielding the solution
\beq\label{eq:h34}
h_3 (x = t-t_2) = \frac1{\l w_0} \left\{ - x + \frac1{\l w_0} \left[ 2 + W_0 \left( - w(z) e^{-w(z)} \right) 
\right] \left( e^{\l w_0 x} -1 \right)  \right\} \ .
\eeq
The exit time at which $h(t) = 1/\l$ is given by
\beq\label{eq:dt34}
\d t_{34}(z) = t_4 - t_3 = - \frac1{\l w_0} \left\{  w_{34}(z) + \l w_0^2 + 
W_{-1} \left[ - w_{34}(z) e^{-\left( w_{34}(z) + \l w_0^2 \right)} \right] \right\} \ ,
\eeq
having called $w_{34}(z) = 2 + W_0 \left( - w(z) e^{-w(z)} \right)$. The secondary branch 
$W_{-1}$ of the Lambert function has been chosen because of the condition $\d t_{34}>0$.

For $t>t_4$, the particle leaves the attractive region and diverges to $h \to \io$ without 
giving any further contribution to the kernels.

\subsection{Fluctuating response}
\label{app:DMFT-H}

Before proceeding with the computation of the kernels, we need to compute the fluctuating 
response $H(t,s)$ in any of the previously defined zones. In the dilute limit (first iteration),
the dynamics of $H(t,s)$ in Eq.~\eqref{eq:H} reduces to (working in rescaled time)
\beq\label{eq:Hdilute}
\frac{\partial}{\partial t} H(t,s) = - \wh V''(h(t)) \left[ H(t,s) - \d (t-s) \right] \ .
\eeq
We know that $H(t,s)=0 \: \: \forall t<s$ because of  causality. The delta term in the rhs 
is equivalent to an initial condition $H(t=s^+,s) = \wh V''(h(s))$. Therefore, Eq.~\eqref{eq:Hdilute}
has the general solution
\beq\label{eq:Hsol}
H(t,s) = 
\begin{cases}
	0 & t < s \\
	\wh V''(h(s)) \exp \left[ - \int_s^t \de t' \wh V''(h(t')) \right] & t > s
\end{cases}
\ .
\eeq
The potential defined in Eq.~\eqref{eq:sticky_pot} has a piece-wise constant second derivative; we can 
compute $H(t,s)$ as a piece-wise defined function depending only on the time zones. Since 
$H(t,s)>0$ only if $s$ is in a region where interaction is present, we can restrict the 
computation to these zones. Furthermore, the definition of $\MM_R(t,s)$ in Eq.~\eqref{eq:ker} 
shows that there is a contribution only at times $t$ where the interaction is present, then 
we will consider only the cases $t_1 < s < t < t_5$ (tangential trajectories) and 
$t_1 < s < t < t_4$ (colliding trajectories).

\subsubsection{Tangential trajectories}

We have only one time zone, so $t_1 < s < t < t_5$. In this region the second derivative is 
constant and has $\wh V''(h) = - \l w_0$, so

\beq\label{eq:H15}
H_{15}(t,s) = - \l w_0 e^{\l w_0 (t-s)} \qquad \qquad t_1<s<t<t_5 \ .
\eeq

\subsubsection{Colliding trajectories}

Following the same reasoning as above and using Eq.~\eqref{eq:Hsol}, we can compute 
$H(t,s)$ for any possible combination of $t_1<s<t<t_4$, which will include ``self'' terms (when 
$s$ and $t$ are in the same time zone) and ``mixed'' terms (when they belong to different 
zones). So, for the self terms we find
\beq\label{eq:H12}
H_{12}(t,s) = - \l w_0 e^{\l w_0 (t-s)} \qquad \qquad t_1<s<t<t_2 \ ,
\eeq
\beq\label{eq:H23}
H_{23}(t,s) = \l w_0 e^{-\l w_0 (t-s)} \qquad \qquad t_2<s<t<t_3 \ ,
\eeq
\beq\label{eq:H34}
H_{34}(t,s) = - \l w_0 e^{\l w_0 (t-s)} \qquad \qquad t_3<s<t<t_4 \ ,
\eeq
and for the mixed terms
\beq\label{eq:H13}
H_{13}(t,s) = - \l w_0 e^{-\l w_0 (t-2t_2+s)} \qquad \qquad t_1<s<t_2<t<t_3 \ ,
\eeq
\beq\label{eq:H14}
H_{14}(t,s) = - \l w_0 e^{\l w_0 (t-2t_3+2t_2-s)} \qquad \qquad t_1<s<t_2<t_3<t<t_4 \ ,
\eeq
\beq\label{eq:H24}
H_{24}(t,s) = \l w_0 e^{\l w_0 (t-2t_3+s)} \qquad \qquad t_2<s<t_3<t<t_4 \ .
\eeq
    
    \subsection{Kernels}
    \label{app:DMFT-ker}
    
    We now compute the dynamical kernels to the first order in the 
    rescaled density $\wh\f$, starting from the definitions given 
    in Eq.~\eqref{eq:ker}.
    
    First, we note that these can be computed as the sum of the kernels  computed separately on 
the different time zones, namely
\beq\label{eq:kappa_total}
\k(t) = \k_{15}(t) + \k_{12}(t) + \k_{23}(t) + \k_{34}(t) \ ,
\eeq
and
\beq\label{eq:MR_a2}
\MM_R(t,s) = \MM_R^{15}(t,s) + \MM_R^{12}(t,s) + \MM_R^{23}(t,s) + \MM_R^{34}(t,s) + 
\MM_R^{13}(t,s) + \MM_R^{14}(t,s) + \MM_R^{24}(t,s) \ .
\eeq
Second, as mentioned in Eq.~\eqref{eq:MR_exp} we assume that the retarded 
memory $\MM_R(t,s)$ is short ranged because of the vanishing duration of a 
collision in the hard-core limit; we are therefore interested in 
computing the stiffness $\g(t)$ and the friction correction $\c_1(t)$,
defined as
\beq
\begin{split}
	\g(t) &= \k(t) - \int_0^t \de s \, \MM_R(t,s) = \k(t) - \c_0(t)\\
	\c_1(t) &= \int_0^t \de s \, \MM_R(t,s) \, (t-s)
\end{split}
\eeq 
We expect that $\g(t)$ vanishes in the steady state, so that $h(t)$ is not 
confined at long times (otherwise we would be in the glassy phase at any 
density) and that $\c_1(t)$ goes to a constant depending on the density, 
giving us the first-order density correction to the activity and to the MSD.

We will show in  Appendix~\ref{app:DMFT-hw} that higher order terms do 
not contribute to the dynamics.

\subsubsection{Change of variables}

We perform the computation of the kernels in the $(x,z)$ plane, 
being $x = t-t_i$ for every time zone starting in $t_i$ and 
$z = \l w_0 \a -1$, as defined in Sec.~\ref{app:DMFT-traj}. 
Since $t_1 = - \x_0 - \a > 0$, then $\x_0 = - \a - t_1 = - (1+z)/(\l w_0) - t_1$.\\
When performing the integrals over $\x_0$ and $z$, we will choose the 
normal region $z>-1$ and $\x_0 < - (1+z)/(\l w_0)$. The latter condition 
implies $t_1>0$.

This choice is particularly convenient to implement the time zone conditions \eg $\Th (t_i < t < t_j)$. The former actually translates to $0<x<\d t_{i,j}$ in every time region, where 
one typically has $j=i+1$. So we move from the integration over $\x_0$ to the integration over 
$x = t- t_i = t - \d t_{1i} - t_1 = t - \d t_{1i} + \x_0 + (1+z)/(\l w_0)$.\\
The condition $\x_0 < - (1+z)/(\l w_0)$ then implies $x < t - \d t_{1i}$. 
The typical times $\d t_{1i}$ are those computed in 
Appendix~\ref{app:DMFT-traj}; but since we are interested in the 
long-time limit, the assumption $t \to \io$ automatically 
satisfies this condition; hence the integration region for $\a > 0$, $\x_0 < 
-\a$ and $t_i < t < t_j$ is equivalent in the long time limit 
to\footnote{If one wants to recover the time dependence of the kernels, it is sufficient 
to substitute the upper bound of the integration over $x$ with $\min(\d t_{ij}(z), t - \d 
t_{1i})(z)$.}
\beq\label{eq:integration-region}
z > -1 \quad , \quad  0 < x < \d t_{ij} (z) \ .
\eeq
For tangential trajectories, we have $-1<z<z_2$, while for colliding trajectories we have 
$z>z_2$.

The Gaussian weight in the integral then becomes
\beq
\frac{\wh\f}2 \frac{e^{1/\l}}{\sqrt{2\p}} \, \a \, e^{-\a^2/2} \, \de \a \, \de \x_0 = \frac{\wh\f}2 
\frac{e^{1/\l}}{\sqrt{2\p}} \, \frac{1+z}{(\l w_0)^2} \, e^{-(1+z)/(2(\l w_0)^2)} \, \de z \, \de x 
\equiv I_0(z) \, \de z \, \de x \ .
\eeq
With all these precautions we can directly plug into the kernel 
integration the trajectories computed as functions of $z,x$ in 
Sec.~\ref{app:DMFT-traj}. The computation is tedious but straightforward, 
and we will repeatedly apply the following formulas:
\beq\label{eq:ker_comp}
\begin{split}
\k_{ij} &= \int \de z \, I_0(z) \int_0^{\d t_{ij}(z)} \de x \, \left[ \wh V''(h(t)) + \wh V'(h(t)) \right] \ , \\
\c_n^{ij} &= \int \de z \, I_0(z) \int_0^{\d t_{ij}(z)} \de x \, \wh 
V''(h(t)) \int_0^t \de s \, H_{ij} (t,s) (t-s)^n \, \th(t_i<s<t) \ , \\
\end{split}
\eeq
where $ij$ are the time zone indices, integrating over the appropriate domain 
of $z$ and recalling that $t=t_i+x$.

\subsubsection{Tangential trajectories}

For tangential trajectories we only have one time zone $t_1<t<t_5$ and the 
tangential condition $-1<z<z_2$: using Eqs.~\eqref{eq:h12_z} 
and~\eqref{eq:H15} we find
\beq\label{eq:Gamma15}
\g_{15} = \int_{-1}^{z_2} \de z \, I_0(z) \left[ - \frac{z}{\l w_0} \d t_{15}(z) + \frac12 \d t^2_{15}(z) + \left( -1 + \frac{z}{(\l w_0)^2} \right) \left( e^{\l w_0 \d t_{15}(z)} - 1 \right) 
\right] \ ,
\eeq
and
\beq\label{eq:c1_15}
	\c_1^{15} = \int_{-1}^{z_2} \de z \, I_0(z) \left[ \d t_{15}(z) + \frac2{\l w_0} + \left( \d t_{15}(z) - \frac2{\l w_0} \right) e^{\l w_0 \d t_{15}(z)} \right] \ .
\eeq

\subsubsection{Colliding trajectories}

We have now several time zones and the collisional condition $z>z_2$.
We explicitly write  the result for every time zone following 
Eq.~\eqref{eq:ker_comp}.

Stiffness terms:
\beq\label{eq:gamma_coll}
\begin{split}
\g_{12} &= \int_{z_2}^{+\io}  \de z \, I_0(z) \left[ - \frac{z}{\l w_0} \d t_{12}(z) + \frac12 \d t^2_{12}(z) + \left( -1 + \frac{z}{(\l w_0)^2} \right) \left( e^{\l w_0 \d t_{12}(z)} - 1 \right) 
\right] \ , \\
\g_{23} &= \int_{z_2}^{+\io}  \de z \, I_0(z) \left[ \left( w_0 - 
\frac{w(z)}{\l w_0} \right) \d t_{23}(z) + \frac12 \d t^2_{23}(z) + \left( 1 + \frac{w(z)}{(\l w_0)^2} \right) \left( 1 - e^{-\l w_0 \d t_{23}(z)} \right) \right] \ , \\
\g_{34} &= \int_{z_2}^{+\io}  \de z \, I_0(z) \left[ \left( w_0  + 
\frac{w_{34}(z)}{\l w_0} \right) \d t_{34}(z) + \frac12 \d t^2_{34}(z) + 
\left( -1 - \frac{w_{34}(z)}{(\l w_0)^2} \right) \left( e^{\l w_0 \d 
t_{34}(z)} - 1 \right) \right] \ , \\
\g_{13} &= \int_{z_2}^{+\io} \de z \, I_0(z) \left( e^{\l w_0 \d t_{12}(z)} -1 \right) \left( 1 - e^{-\l w_0 \d t_{23}(z)} \right) \ , \\
\g_{14} &= - \int_{z_2}^{+\io} \de z \, I_0(z) \, e^{-\l w_0 \d t_{23}(z)} \left( e^{\l w_0 \d t_{34}(z)} -1 \right) \left( e^{\l w_0 \d t_{12}(z)} -1 \right) \ , \\
\g_{24} &= \int_{z_2}^{+\io} \de z \, I_0(z) \left( e^{\l w_0 \d t_{34}(z)} -1 \right) \left( 1 - e^{-\l w_0 \d t_{23}(z)} \right) \ .
\end{split}
\eeq
Friction correction:
\beq\label{eq:c_1_coll}
\begin{split}
	\c_1^{12} &= \int_{z_2}^{+\io}  \de z \, I_0(z) \left[ \d t_{12}(z) + \frac2{\l w_0} + \left( \d t_{12}(z) - \frac2{\l w_0} \right) e^{\l w_0 \d t_{12}(z)} \right] \ , \\
	\c_1^{23} &= \int_{z_2}^{+\io}  \de z \, I_0(z) \left[ \d t_{23}(z) - \frac2{\l w_0} + \left( \d t_{23}(z) + \frac2{\l w_0} \right) e^{-\l w_0 \d t_{23}(z)} \right] \ , \\
	\c_1^{34} &= \int_{z_2}^{+\io}  \de z \, I_0(z) \left[ \d t_{34}(z) + \frac2{\l w_0} + \left( \d t_{34}(z) - \frac2{\l w_0} \right) e^{\l w_0 \d t_{34}(z)} \right] \ , \\
	\c_1^{13} &= \int_{z_2}^{+\io} \de z \, I_0(z) \left[ -\d t_{12} (z) e^{\l w_0 \d t_{12}(z)} - \d t_{23}(z) e^{-\l w_0 \d t_{23}(z)} + \d t_{13}(z) e^{-\l w_0 [\d t_{23}(z) - \d t_{12}(z) ]} \right] \ , \\
	\c_1^{14} &= \int_{z_2}^{+\io} \de z \, I_0(z) \, e^{-\l w_0 \d t_{23}(z)} \left[ -\frac2{\l w_0} + \d t_{23}(z) +
	\left( \frac2{\l w_0} - \d t_{13}(z) \right) e^{\l w_0 \d t_{12}(z)} + \left( \frac2{\l w_0} - \d t_{24}(z) \right) e^{\l w_0 \d t_{34}(z)} + \right. \\
	& \qquad \qquad \qquad \qquad \qquad \qquad \left. + \left( - \frac2{\l w_0} + \d t_{14}(z) \right) e^{\l w_0 \left( \d t_{12}(z) + \d t_{34}(z) \right)} \right] \ , \\
	\c_1^{24} &= \int_{z_2}^{+\io} \de z \, I_0(z) \left[ -\d t_{23}(z) e^{-\l w_0 \d t_{23}(z)}  -\d t_{34} (z)
	e^{\l w_0 \d t_{34}(z)} + \d t_{24}(z) e^{\l w_0 [\d t_{34}(z) - \d t_{23}(z) ]} \right] \ .
\end{split}
\eeq

\subsection{Hard-sphere limit}
\label{app:DMFT-hw}

The expressions written above are \textit{exact} in the long-time limit.
To obtain an analytical expression, we move to the hard-sphere limit 
$\l \to\io$, which we use to approximate the behavior of the 
stiffness and of the friction correction.

The analytical computation requires the approximation of the Lambert  
function $W$  in the different intervals of the integration over $z$. 
We computed the integrations in the previous equations both analytically
and numerically; we omit the details of the computation because they are 
tedious. Altogether, the only terms 
that survive when $\l \to\io$ are
\begin{gather}
    \g_{15} = - \g_{23} = - w_0^2/2 \ , \\
    \c_1^{15} = \frac{\wh\f}{6\sqrt{2\p}} w_0^3 \ , \quad 
    \c_1^{23} = \frac{\wh\f}4 \ .
\end{gather}
This final result is crucial and tells us that (i) the elastic response 
$\g(t)$ vanishes in the long-time limit, and the particles can diffuse;
(ii) the friction correction $\c_1$ leading to the effective 
self-propulsion has two contributions, one coming from the purely repulsive
interaction ---$\c_1^{23}$--- and the other from the attractive region 
---$\c_1^{15}$--- , so that one finally finds

\beq
\c_1 = \frac{\wh\f}4 \left( 1 + \frac{\sqrt2}{3\sqrt\p} w_0^3 \right) \ .
\eeq

\subsubsection{Vanishing terms}

Here we sketch the reason why we stopped to the first-order in the 
expansion of the integrated response in Eq.~\eqref{eq:MR_exp}: 
when we need to compute the integral 
$\int^t_{t_i} \de s \, H(t,s) (t-s)^n$, the fluctuating response 
has an exponential behavior and decays with a characteristic time 
$(\l w_0)^{-1}$. Therefore, when computing the instantaneous response
$\k$ and the zero-th order contribution $\c_0$, these both diverge
separately as $\OO(\l)$ in the hard-sphere limit but their difference 
has a finite limit. When computing $\c_1$, the first degree term $(t-s)$ 
in the integral lowers one degree in $\l$ and its contribution is 
therefore finite. 

This scheme repeats when computing $\c_2$, and lowering another
degree in $\l$ implies $\c_2 = \OO(\l^{-1})$, therefore all 
the $\c_n$ vanish in the hard-sphere limit for $n\geq2$. 

This argument can be carried out with explicit analytical results 
in the dilute phase where one has a specific solution for $H(t,s)$.
However, given that its validity is provided by the hard-wall limit 
of the interactions yielding a vanishing relaxation time of the response 
kernels, we conjecture that the same argument is valid at any density.

\end{document}